\documentclass[intlimits,twoside,a4paper]{article}
\usepackage{graphicx}
\usepackage{wrapfig}
\usepackage[cp1251]{inputenc}
\usepackage{gensymb}
\usepackage{multirow}
\usepackage{afterpage}

\usepackage{cmpj3}

\issue{2017}{20}{2}{23002}
\doinumber{10.5488/CMP.20.23002}

\title[Membrane separation for methane-hydrogen gas mixtures]%
{Membrane separation study for methane-hydrogen gas mixtures by molecular simulations}

\author[T. Kov\'acs, S. Papp, T. Krist\'of]{T. Kov\'acs, S. Papp, T. Krist\'of\footnote{Author for correspondence, E-mail:~kristoft@almos.vein.hu.}}
\address{Department of Physical Chemistry, University of Pannonia, P. O. Box 158, Veszpr\'em, H-8201, Hungary
 }
\date{Received March 16, 2017, in final form April 19, 2017}

\begin{document}

\maketitle

\begin{abstract}

Direct simulation results for stationary gas transport through pure silica zeolite membranes (MFI, LTA and DDR types) are presented using a hybrid, non-equilibrium molecular dynamics simulation methodology introduced recently.
The intermolecular potential models for the investigated CH$_{4}$ and H$_{2}$ gases were taken from literature.
For different zeolites, the same atomic (Si and O) interaction parameters were used, and the membranes were constructed according to their real (MFI, LTA, or DDR) crystal structures.
A realistic nature of the applied potential parameters was tested by performing equilibrium adsorption simulations and by comparing the calculated results with the data of experimental adsorption isotherms.
The results of transport simulations carried out at 25\degree{C} and 125\degree{C}, and at 2.5, 5 or 10~bar clearly show that the permeation selectivities of CH$_{4}$ are higher than the corresponding permeability ratios of pure components, and significantly differ from the equilibrium selectivities in mixture adsorptions.
We experienced a transport selectivity in favor of CH$_{4}$ in only one case.
A large discrepancy between different types of selectivity data can be attributed to dissimilar mobilities of the components in a membrane, their dependence on the loading of a membrane, and the unlike adsorption preferences of the gas molecules.

\pacs 02.70.Ns, 07.05.Tp, 68.43.-h, 68.43.Jk
\end{abstract}

\section{Introduction}
\label{sec:intro}

Zeolites are made up of silicon, aluminum, and oxygen atoms linked together so that they form structurally well-defined pores.
A high regularity of the structure distinguishes zeolites from other porous materials and makes their high selectivity in catalytic and separation processes possible \cite{1,2}.
A usual size of zeolite micropores is similar to that of many small molecules.
Therefore, it is possible for some molecules to enter the zeolite and then get stuck in the pores, where they can even react with each other, while the other molecules can move through the zeolite channels faster.
In separation processes, the advantage of zeolites over other types of adsorbents/membranes is that they typically offer good endurance to high pressure and temperature and can often tolerate harsh chemical environments.
The Si-Al ratio of the zeolite framework is an important factor of the application.
Zeolites with lower Si-Al ratios are more hydrophilic, while zeolite membranes with higher Si-Al ratios have fewer
structural defects.
High-silica zeolite membranes are greatly preferred in gas separation applications, and among them silicalite (MFI) is one of the most commonly studied.

Purification of methane from carbon dioxide (carbon dioxide is one of the major contaminants of natural gas) turns out to be an especially attractive gas separation process.
Another important practical system that is involved in the process of purification of synthetic gas obtained from steam reforming of natural gas is the methane-hydrogen system.
The available literature data, however, are scarce for all the systems that combine methane-hydrogen gas mixtures with the use of zeolite membranes \cite{3,4}.

The behavior of material systems, characterized at the atomic level, can be effectively studied using classical molecular simulations. The atomistic simulation results can explain or, in some cases, substitute the experimental results.
Nowadays, these simulations play an important role in the description of processes that occur in crystalline adsorbents and membranes \cite{5,6,7,8,9,10,11,12,13}.
To simulate steady-state transport of molecules, we need a method that satisfies two criteria: it must ensure the real dynamics of a system, and it must maintain the steady-state driving force at the microscopic level.
The dynamics of a system can be investigated microscopically by means of molecular dynamics (MD), dynamic Monte Carlo \cite{8} and other direct or indirect simulation methods (e.g., a very fast NP+LEMC method \cite{11}).
To preserve a constant driving force, these methods are often linked with the other techniques, e.g., dual control volume or local equilibrium Monte Carlo (LEMC) techniques \cite{12}.
There are many composite methods having many limits of applicability and numerous advantages and disadvantages, such as the gradient relaxation MD \cite{13} or other earlier gradient techniques (e.g., \cite{14,15}), a self-adjusting plates technique \cite{16}, the external field MD \cite{17} or its boundary-driven version \cite{18}, and the dual control volume grand canonical molecular dynamics (DCV-GCMD) \cite{5}.
In general, the main problem with simulating steady-state membrane transport at the atomic level is that we cannot simulate (at least virtually) bulk fluid phases of realistic size.
Therefore, particle depletion/accumulation in the simulated bulk regions frequently occurs, if particle reinjection/removal steps are not applied.
A sudden appearance and annihilation of molecules, however, can disturb the steady-state flux of transporting particles.
Recently, we developed a simple atomistic simulation method for membrane transport that can maintain a driving force without significantly disturbing the previously-developed steady-state flux \cite{19}, while it properly mimics the common experimental situation in gas permeations measurements, where pressure is the main control parameter.

In this work, direct simulation results for steady-state gas transport through some relevant pure silica zeolite membranes are reported using our novel hybrid MD simulation method. In what follows, we briefly outline the applied transport simulation technique, and then  specifications of the performed simulations and results for the separation of the technologically important CH$_{4}$-H$_{2}$ mixture (that might also be relevant in the development of engine fuels with high hydrogen content) will be presented in more detail.

\section{Transport simulation method}
\label{sec:method}

The idea behind our method is based on the fact that in experimental gas permeation arrangements, pressure is the property that can be controlled relatively simply: the partial pressure for each component on the input side of the membrane and the total pressure on the permeate side.
The essential point of our atomistic simulation scheme (called pressure-tuned, boundary driven molecular dynamics technique, PBD-MD) is that  pressure is controlled by adjusting the particle numbers on the two sides of the membrane indirectly \cite{19}.

We started from the DCV-GCMD approach \cite{5} containing two equilibrium control cells distinguished by unequal intensive thermodynamic parameters (such as chemical potential or pressure) on the two sides of the membrane, where random particle insertion/deletion moves are applied to maintain bulk phase properties in the control cells and thus to uphold the desired (constant) driving force through the membrane.
However, to avoid such non-physical particle moves in the vicinity of the membrane affecting any eventual sorption layers on the membrane surfaces and interfering with earlier stabilized flux of the permeating particles, adjusting the particle numbers is restricted to zones far from the membrane region.
This means that while all properties calculated and monitored in the control cells predominantly stem from the movements and interactions of particles in these regions, the artificial perturbation of the system is only present in the boundary regions of the simulation cell \cite{18}.
Our earlier test clearly showed that the particles inserted into the system have no ``memories'' of their initial velocities before they reach the interaction range of the membrane \cite{19}.

Here, we consider the system as being at a constant temperature $T$ with a fixed number of particles $N$ and a box volume $V$.
To attain the target pressure $p$ (or partial pressures in the case of mixtures), regular perturbations in the number of particles are applied close to the boundaries of the total simulation box.
In this way, as the chemical potentials of the control cells used in the traditional DCV-GCMD technique are connected to the corresponding partial pressures, the pressure can be controlled effectively in the control cells.
In our trial-and-error type pressure-tuning approach, the particle insertion and deletion steps are performed randomly (i.e., randomly chosen particles and positions) at both ends of the simulation box, in the direction of the transport, far enough from the membrane region (in this respect, this technique is similar to the Particle Counting (PACO) method of Berti et al. \cite{20}).
One particle insertion ($+1$ case) or deletion ($-1$ case) is executed periodically if the following general criterion is satisfied:
\begin{equation}
\left| \dfrac{N_{\mathrm{control}\,\mathrm{cell}}\pm 1}{N_{\mathrm{control}\,\mathrm{cell}}} p_{\mathrm{control}\,\mathrm{cell}} - p_{\mathrm{target}} \right| < \left| p_{\mathrm{control}\,\mathrm{cell}} - p_{\mathrm{target}} \right|,
\end{equation}
where $N_{\mathrm{control}\,\mathrm{cell}}$ is the number of particles in the control cell, and $p_{\mathrm{control}\,\mathrm{cell}}$ and $p_{\mathrm{target}}$ are the calculated pressure and the designated pressure, respectively, in the same control cell.
Here, $p$ denotes partial pressure on the feed side or total pressure on the permeate side, and $N_{\mathrm{control}\,\mathrm{cell}}$ denotes the number of particles of the individual components on the feed side or the total number of particles on the permeate side.
The attempts are always accepted, except for those random insertion steps which result in particle overlap (in such cases, the insertion attempt is repeated).

\section{Simulation details}
\label{sec:sim}

\begin{figure}[!b]
\begin{center}
\scalebox{0.75}{\includegraphics{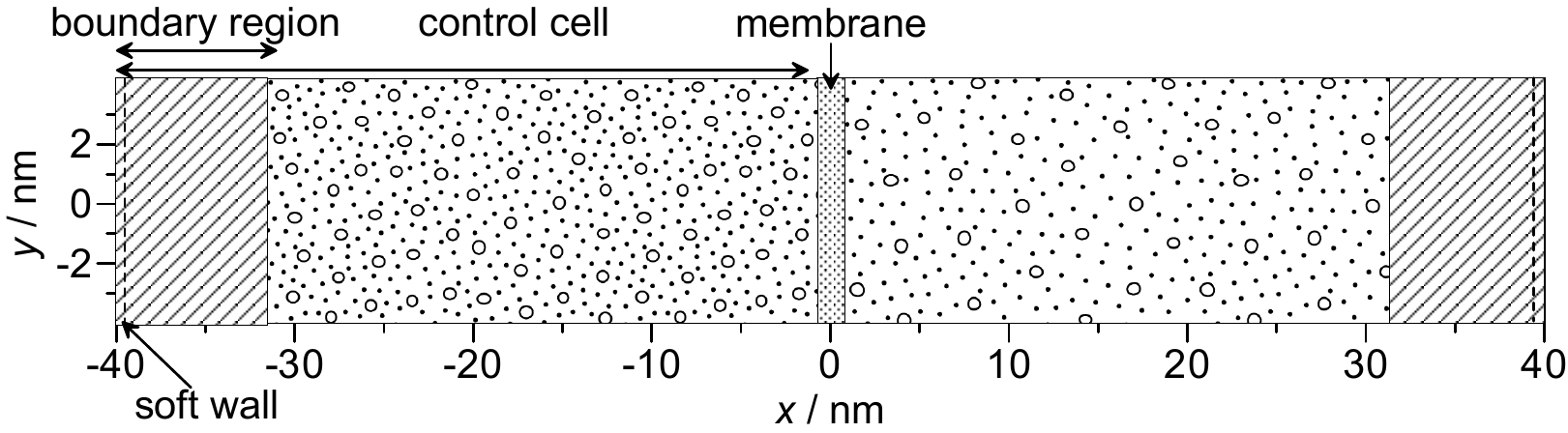}}
\end{center}
\caption{The simulation box used in the PBD-MD method.}
\label{Fig1}
\end{figure}

In the transport simulations, standard MD was used employing the leap-frog algorithm as an integrator with the time step of 2~fs. The Berendsen thermostat with the thermal coupling parameter of 200~fs (very weak coupling) was applied for controlling the temperature.
The temperature was defined by subtracting off the streaming velocity from the $x$ component of the particle velocities, where $+x$ direction is the direction of the transport.
The simulation box was confined by impenetrable, soft repulsive walls in the direction of the transport, while periodic boundary conditions were applied in the other two directions. (Note that the use of periodic boundary conditions in the direction of the transport is incompatible with the applied particle number adjustment procedure.)
The geometry of this simulation box is depicted in figure~\ref{Fig1}.
The boundary regions were considerably wider than the range of the repulsive walls on the two sides of the box.
Random initial velocities were assigned to the inserted particles according to the Maxwell-Boltzmann distribution at the prescribed temperature.
One particle insertion or deletion step was performed periodically in both boundary (wall) regions, outside the range of the repulsive walls.
The length of the simulation period between two particle number modifications was taken as 2000 consecutive MD steps and $N_{\mathrm{control}\,\mathrm{cell}}$ and $p_{\mathrm{control}\,\mathrm{cell}}$ were collected as averages for these periods.
The average pressure values were calculated by the virial expression.
The membrane transport processes were simulated for at least 30 million MD steps, but in several cases twice as long runs were needed to collect a sufficient amount of transfer events (at least several hundred).

We also executed equilibrium adsorption simulations in the grand canonical ensemble (fixed chemical potential, volume and temperature) using the standard grand canonical Monte Carlo (GCMC) technique.
The pressure/partial pressure of the adsorbate molecules in the gas phase was given indirectly by specifying the chemical potential of the component.
The simulation box size was equal to the size of the investigated zeolite crystal and periodic boundary conditions were applied in all three spatial directions. The external surface adsorption, therefore, was not taken into account.
The adsorption simulations were conducted for 200 million Monte Carlo steps.

\begin{figure}[!b]
\begin{center}
\scalebox{1.4}{\includegraphics{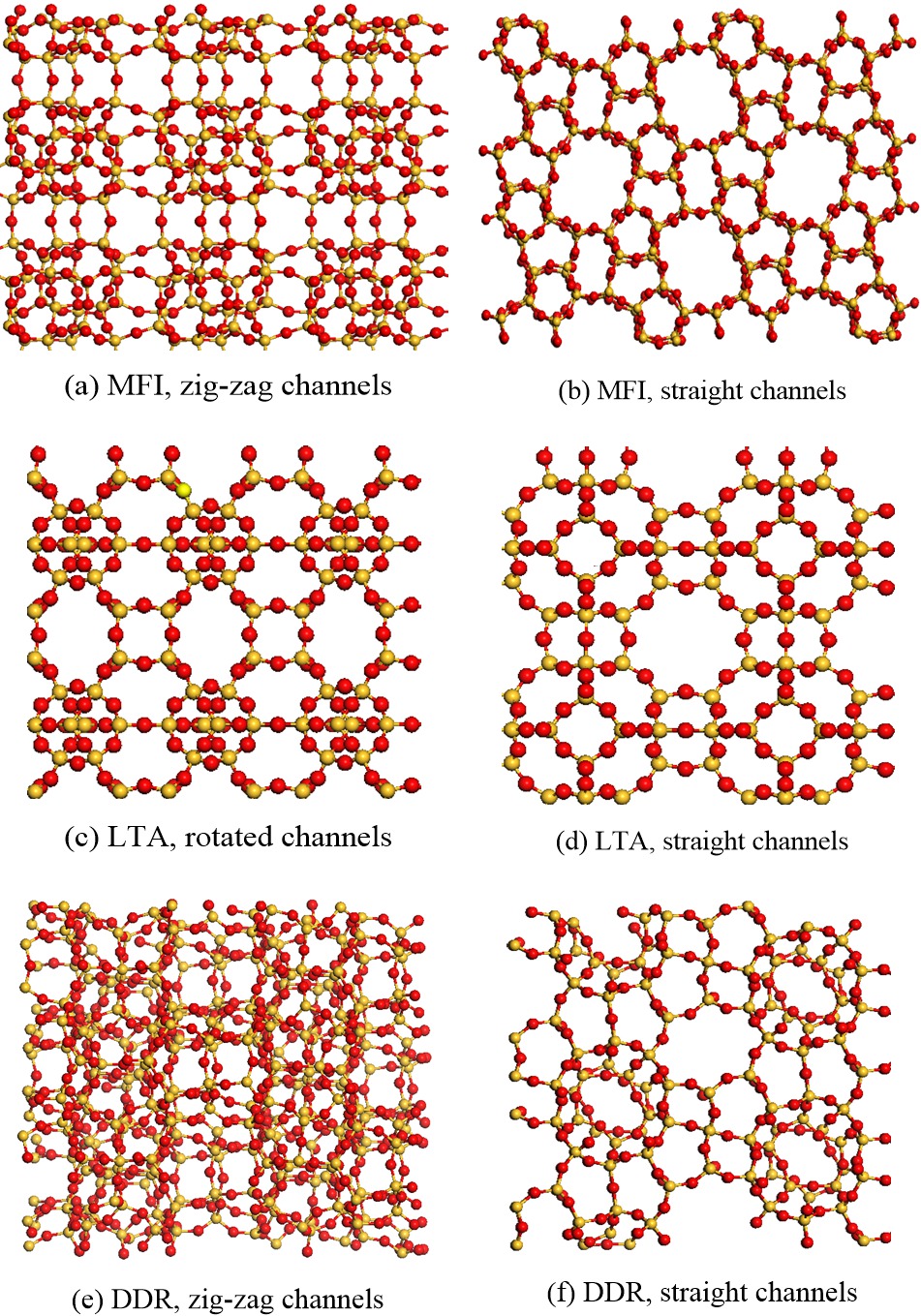}}
\end{center}
\caption{(Color online) Schematic view of the MFI [(a), (b)], LTA [(c), (d)] and DDR [(e), (f)] membranes in the direction of the applied thermodynamic gradient in the transport simulations.}
\label{Fig2}
\end{figure}

We present results for the adsorption (equilibrium state) and transport (steady-state) of gases on all-silica MFI, LTA and DDR zeolites at $T = 298.15$ and 398.15~K, and at $p = 250$ and 500 (or 1000)~kPa.
These are built up from SiO$_{4}$ tetrahedrons, with linear and/or zig-zag (sinusoidal) channels \cite{21}.
In the MFI zeolite [figure~\ref{Fig2}~(a), (b)], straight channels of elliptical cross-section extending along one direction are cross-linked by zig-zag channels of nearly circular cross-section extending in the other perpendicular direction (in both cases the channel diameters are about 0.7~nm).
The crystal lattice parameters of its orthorhombic unit cell are $a = 2.0090$~nm, $b = 1.9738$~nm and $c = 1.3142$~nm, with all the lattice angles being 90\degree.
The pure silicon type of the LTA zeolite [figure~\ref{Fig2}~(c), (d)] has a cubic structure (all the lattice angles are 90\degree) with nearly spherical sodalite cages and straight channels with a diameter of about 1.1~nm between them.
The crystal lattice parameter of its unit cell is $a = 1.1919$~nm.
The pure silica DDR zeolite [figure~\ref{Fig2}~(e), (f)] has a two-dimensional pore network with channels of slightly elliptical cross-section and with the characteristic channel diameter of about 0.8~nm.
The crystal lattice parameters of its hexagonal unit cell are $a = 1.3795$~nm, $b = 1.3795$~nm, $c = 4.0750$~nm, $\alpha =\beta = 90$\degree\ and $\gamma=120$\degree.

The membrane structures were constructed according to the available crystallographic information from the IZA database \cite{21} and taken to be defect-free.
For the adsorption simulations, we built the model structures in the $1\times1\times2$, $2\times2\times2$ and $2\sin\gamma\times2\times1$ arrangements of the unit cells with MFI, LTA and DDR adsorbents, respectively.
In the transport simulations, 1 or 2 unit-cell thick and 10--40~unit-cell wide (tall) membranes were used with the intention that the membrane sizes of the different zeolites should be as similar as possible.
Two crystal orientations were realized in the membranes to place the (1) straight or (2) zig-zag channels of the zeolites in parallel with the $x$ direction [in LTA membranes the straight channels were rotated by 45\degree\ for case (2)].
Thicknesses and surface sizes of these membranes containing $5{-}8000$ Si and O atoms are summarized in table~\ref{tab1}.
The particles of the zeolite lattices were fixed at the crystallographic positions, so the frameworks were kept rigid \cite{22}.
In the steady-state transport simulations, the box size in the direction of the transport was set to 80~nm.

\begin{table}[!h]
\centering
\caption{Thicknesses and surface sizes of the membranes used in the transport simulations.}
\label{tab1}
\vspace{2ex}
\begin{tabular}{|cc|c|c|}
\hline \hline
    &  & zig-zag (or rotated) channels & straight channels  \\ \hline\hline
 \multirow{2}{*}{MFI} & thickness/nm & 2.4200 &  2.2300 \\
                   & surface size/nm$^{2}$ & 41.5035 &  42.2436  \\ \hline
\multirow{2}{*}{LTA} & thickness/nm & 2.6440 & 2.6351   \\
                   & surface size/nm$^{2}$ & 49.6903 &  50.2267  \\ \hline
\multirow{2}{*}{DDR} & thickness/nm & 2.6494 & 3.0200  \\
                   & surface size/nm$^{2}$ & 44.9719 &  48.6832  \\ \hline \hline
\end{tabular}
\end{table}

To describe the molecular interactions, the shifted and cut Lennard-Jones (LJ) pair potential was employed together with the Lorentz-Berthelot combining rule. The effective interaction potential for a pair of particles ($\alpha$ and $\beta$) was calculated as
\begin{equation}
 u^{\alpha\beta}(r)
=\left\{
        \begin{array}{ll}
    u_{\mathrm{LJ}}^{\alpha\beta}(r) - u_{\mathrm{c}}^{\alpha\beta} & \; \mbox{for} \; \;  r< r_{\mathrm{c}}^{\alpha\beta},\\
        0 & \; \mbox{for} \; \; r\geqslant r_{\mathrm{c}}^{\alpha\beta},
        \end{array}
        \right.
\label{eq:pm}
\end{equation}
where
\begin{equation}
u_{\mathrm{LJ}}^{\alpha\beta}(r) = 4\epsilon^{\alpha\beta} \left[ \left( \dfrac{\sigma^{\alpha\beta}}{r} \right)^{12} -
\left( \dfrac{\sigma^{\alpha\beta}}{r} \right)^{6}
\right]
\end{equation}
is the standard 12-6 LJ potential, $u_{\mathrm{c}}^{\alpha\beta}$ is the value of the potential at the cut-off distance $r_{\mathrm{c}}^{\alpha\beta}$, and $\sigma$ and $\epsilon$ are the size and energy parameters of the LJ potential, respectively.
The spherical cut-off distance was set to $r_{\mathrm{c}}^{\alpha\beta}=3.5\sigma^{\alpha\beta}$ leading to $u_{\mathrm{c}}^{\alpha\beta} = -0.00217478 \epsilon^{\alpha\beta}$.
All particles interacted with the walls according to the Weeks-Chandler-Anderson potential.
This is also a shifted and cut LJ potential but it has a short cut-off $2^{1/6}\sigma^{\mathrm{wall}}$, and thus its interaction can only be repulsive. This particle-wall potential was chosen because it is convenient to be used in MD simulations. Its choice does not influence the results in the membrane region.

Single-site (i.e., not atomically detailed) models were used for the gas molecules, and the interaction parameters were taken from the literature \cite{23,24}.
In addition, the same atomic interaction parameters were applied for different zeolites (see table~\ref{tab2}): literature $\sigma$ and $\epsilon$ parameters \cite{23,25} of the zeolitic Si and O atoms were slightly adjusted to better reproduce the available equilibrium adsorption data for MFI.

\section{Results}
\label{sec:res}

In the adsorption simulations, the equilibrium selectivity was defined as follows:
\begin{equation}
S_{\mathrm{E}} = \dfrac{q_{\mathrm{CH}_{4}}}{q_{\mathrm{H}_{2}}}\,.
\end{equation}
Here, $q$ denotes the loading of the zeolite (adsorption inside the adsorbent in $\text{mol}\cdot\text{kg}^{-1}$, relative to the mass of the adsorbent).
In the transport simulations, the permeation selectivity (or dynamical selectivity) was calculated from
\begin{equation}
S_{\mathrm{P}} = \dfrac{j_{\mathrm{CH}_{4}}}{j_{\mathrm{H}_{2}}}\,,
\end{equation}
where $j$ is the component flux.
For comparison, the permeation ratio ($R_{\mathrm{P}}$) was also evaluated; $R_{\mathrm{P}}$ is formally similar to $S_{\mathrm{P}}$ but here the pure component fluxes are used, and thus it can be regarded as the idealized limiting case.

\begin{table}[!t]
\vspace{-1mm}
\centering
\caption{Effective pair potential parameters (size and energy) used in the simulations.}
\label{tab2}
\vspace{2ex}
\begin{tabular}{|c|c|c|}
 \hline \hline
   Atom/molecule & $\sigma$ / nm & $(\epsilon/k_{\mathrm{B}}$) / K  \\ \hline\hline
 O \cite{23} & 0.270 &  130.0 \\
Si \cite{25} & 0.070 &  20.0 \\
CH$_{4}$ \cite{23} & 0.373 &  147.9 \\
H$_{2}$ \cite{24} & 0.296 &  36.7 \\
Soft repulsive wall & 0.300 & 120.0\\ \hline \hline
\end{tabular}
\vspace{-2mm}
\end{table}

First, the applied size and energy parameters of the potential models were tested with equilibrium adsorption calculations for CH$_{4}$ and H$_{2}$.
Single-component adsorption isotherms were determined by GCMC simulations to select/verify the potential parameters for the pure gas components and the zeolite atoms.
Here, we present the results obtained with the final parameters shown in table~\ref{tab2}.

\begin{figure}[!t]
\begin{center}
\scalebox{0.69}{\includegraphics{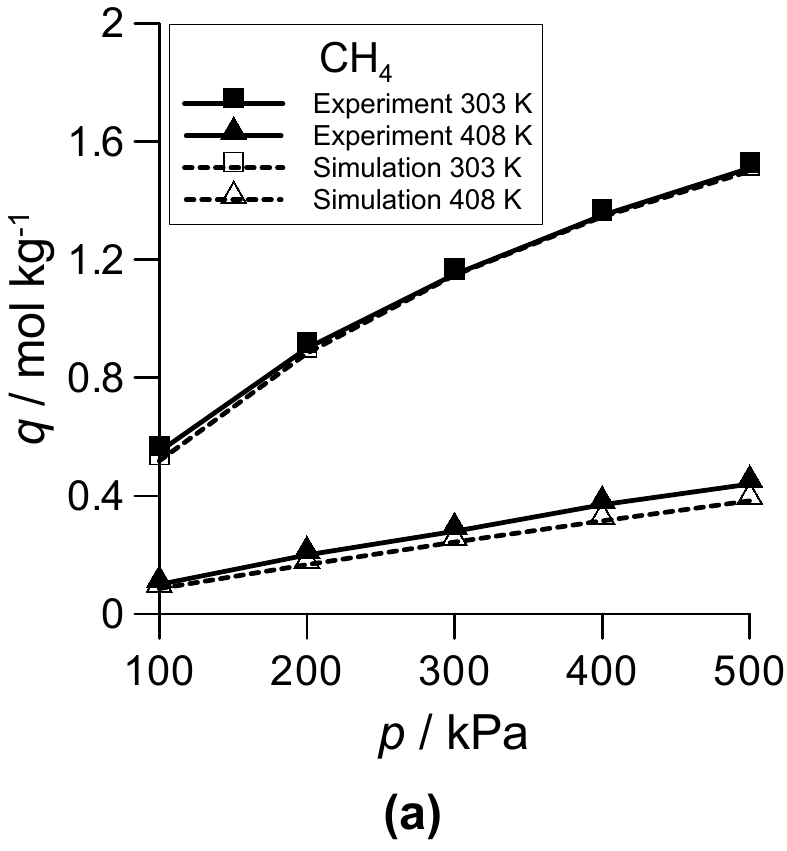}}\scalebox{0.69}{\includegraphics{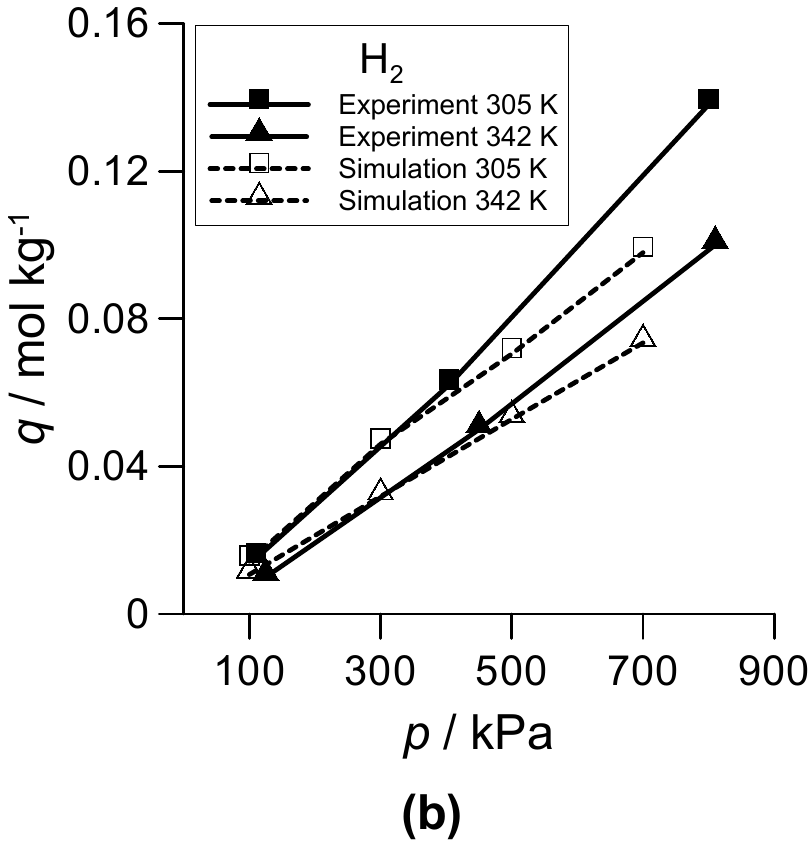}}\\
\scalebox{0.69}{\includegraphics{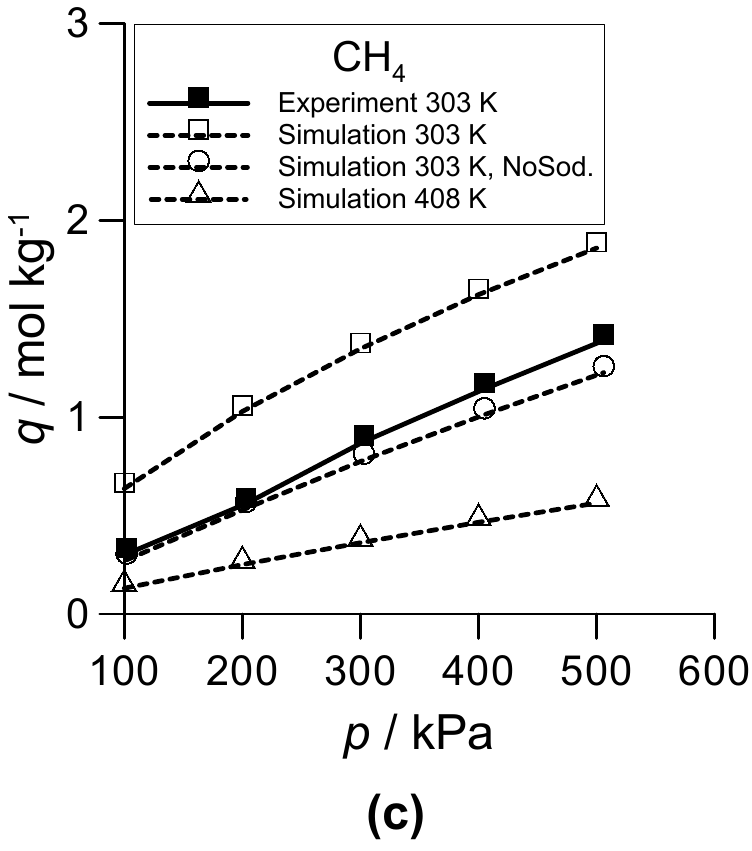}}\scalebox{0.69}{\includegraphics{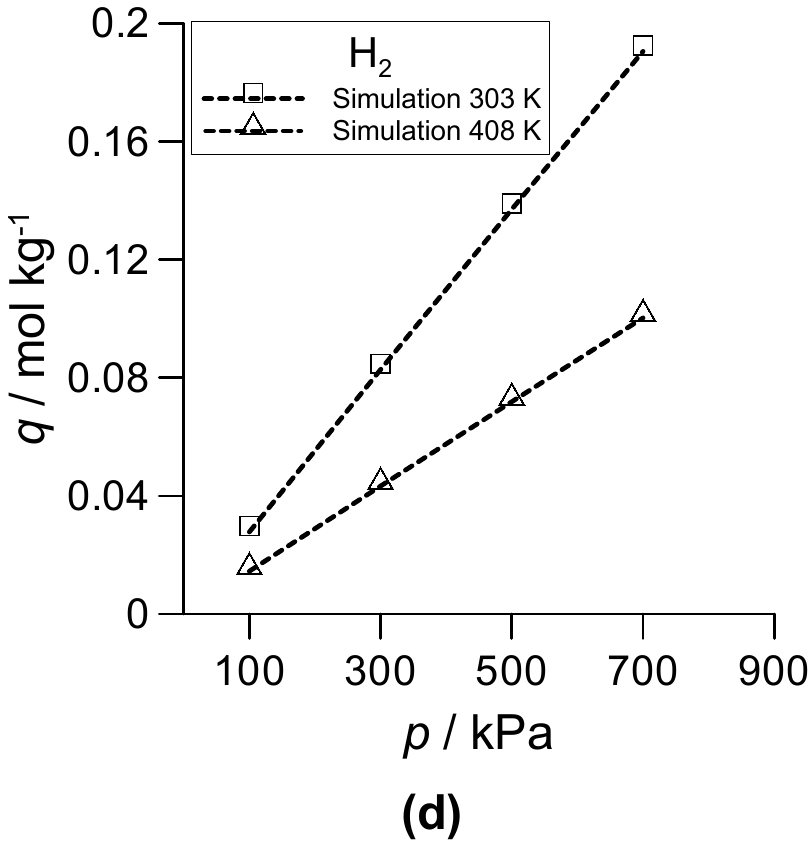}}\\
\scalebox{0.69}{\includegraphics{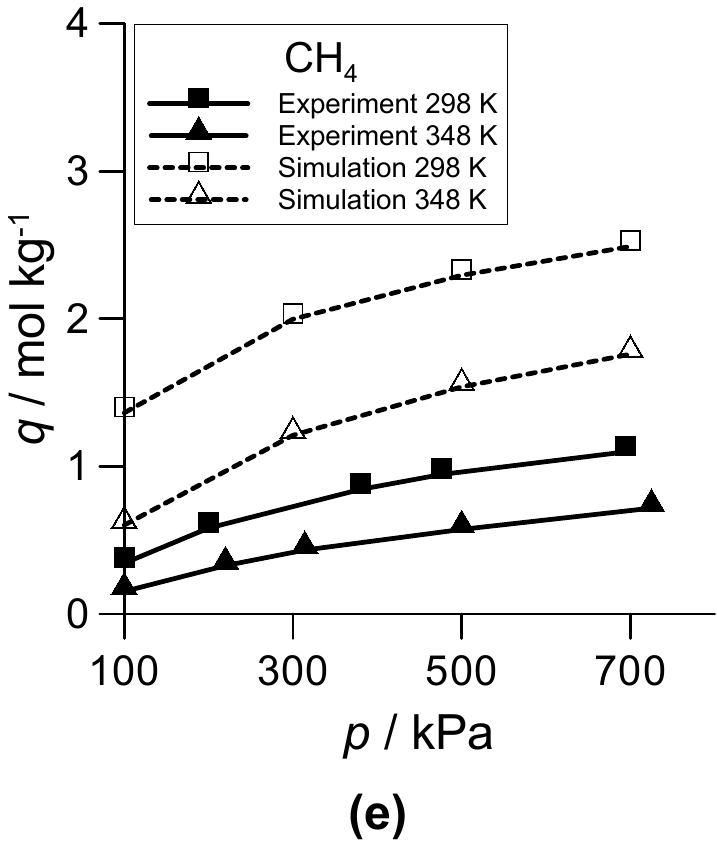}}\scalebox{0.69}{\includegraphics{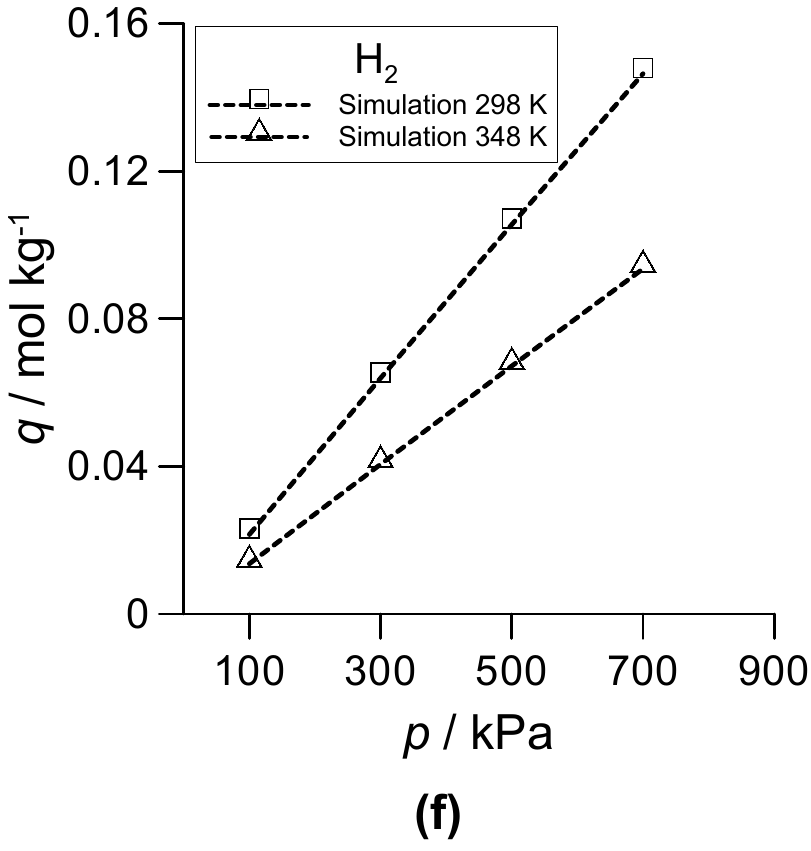}}
\end{center}
\vspace{-3mm}
\caption{Adsorption isotherms of CH$_{4}$ and H$_{2}$ on MFI [(a), (b)], LTA [(c), (d)] and DDR [(e), (f)] zeolites (NoSod.: simulations with the exclusion of sodalite cages). }
\label{Fig3}
\end{figure}

In the case of MFI, as expected, the calculated loadings with CH$_{4}$ agree quite well with the experimental adsorption isotherm data \cite{26} in the investigated temperature and pressure ranges [figure~\ref{Fig3}~(a)].
At lower pressures, a relatively good agreement with the available experimental data \cite{27} can be also observed with H$_{2}$ [figure~\ref{Fig3}~(b)].
There are some inaccessible cavities in this zeolite for the gas molecules, which turned out to be effectively inaccessible in the equilibrium simulations.
For the pure silica LTA zeolite, we found only one experimental adsorption isotherm at room temperature or above, and only with CH$_{4}$, and its data \cite{28} were systematically overestimated in the simulations [figure~\ref{Fig3}~(c), (d)].
However, when we artificially prevented the creations of CH$_{4}$ molecules inside the sodalite cages, an acceptable matching between simulation and measurement could be detected.
In such a way we took into account the physical diffusion pathways in the zeolite, since the CH$_{4}$ molecules are incapable of passing through the small windows of the sodalite cage.
For the pure silica DDR zeolite, experimental data with CH$_{4}$ were available at two temperatures \cite{29} [figure~\ref{Fig3}~(e), (f)] and we did not find such data with pure H$_{2}$.
As it is seen, the simulated adsorption isotherms with CH$_{4}$ deviate considerably from their experimental counterparts.
In our grand canonical simulations, however, this zeolite contained a large quantity of occupiable pores or cavities, which have, according to the literature \cite{30,31}, too small windows to allow molecule access, and so these simulations also overestimate the proper results (in these simulations, due to the complicated geometry, we could not preselect the regions where artificial molecule creations are forbidden).
Accessible volume data from the literature make possible a rough estimation of an average (temperature, pressure and composition-independent) correction factor for the calculated adsorption loading of DDR: this amounts to 0.6 or 0.7 with CH$_{4}$ or H$_{2}$, respectively.
Notwithstanding all these, we must acknowledge that we had to make a compromise at this point to keep the transferability of the pair potential parameters and highlight the differences in the simulated results that originate from the structural alterations between the zeolites.

\begin{figure}[!t]
\begin{center}
\scalebox{0.31}{\includegraphics{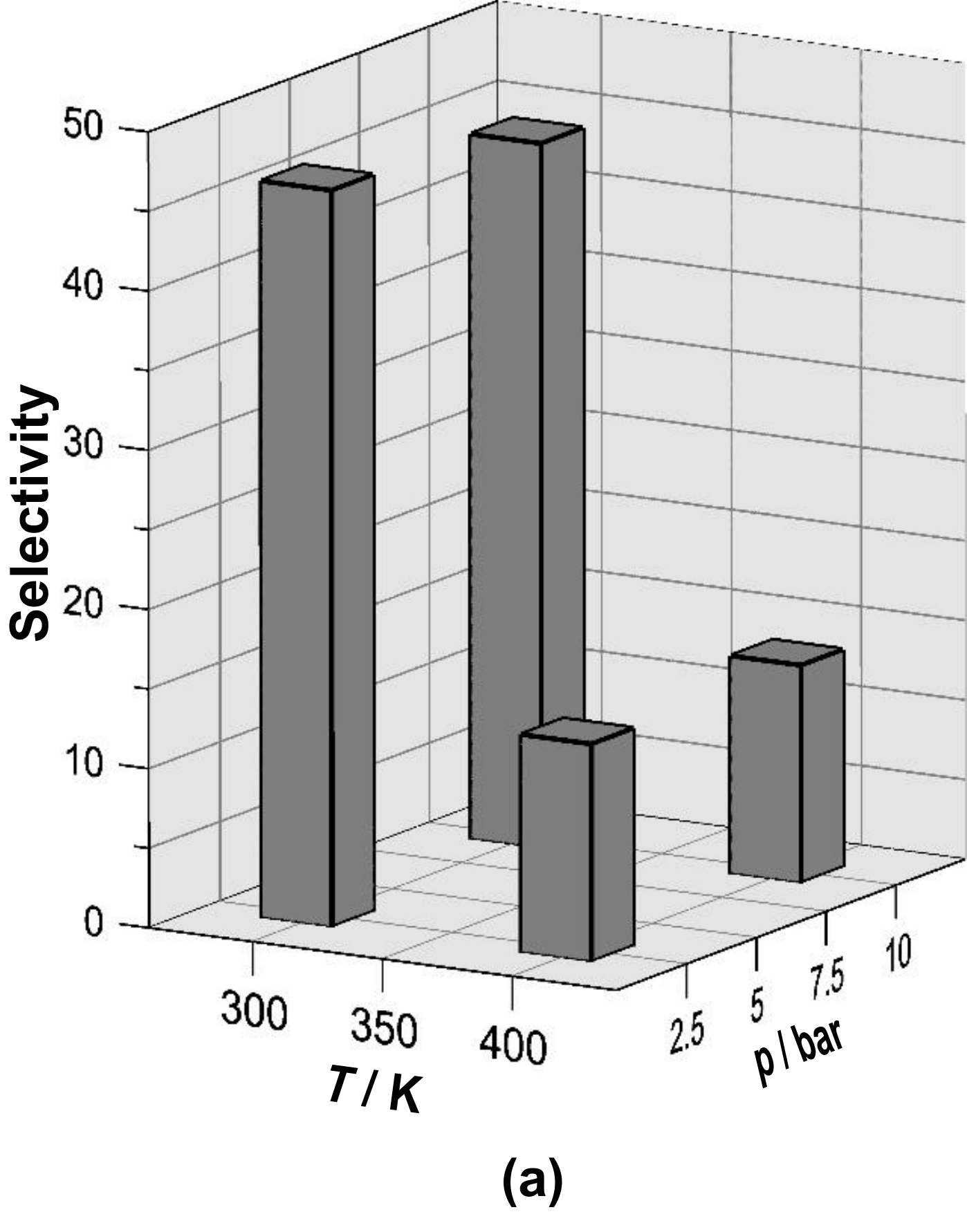}}\hspace{0.5cm}\scalebox{0.31}{\includegraphics{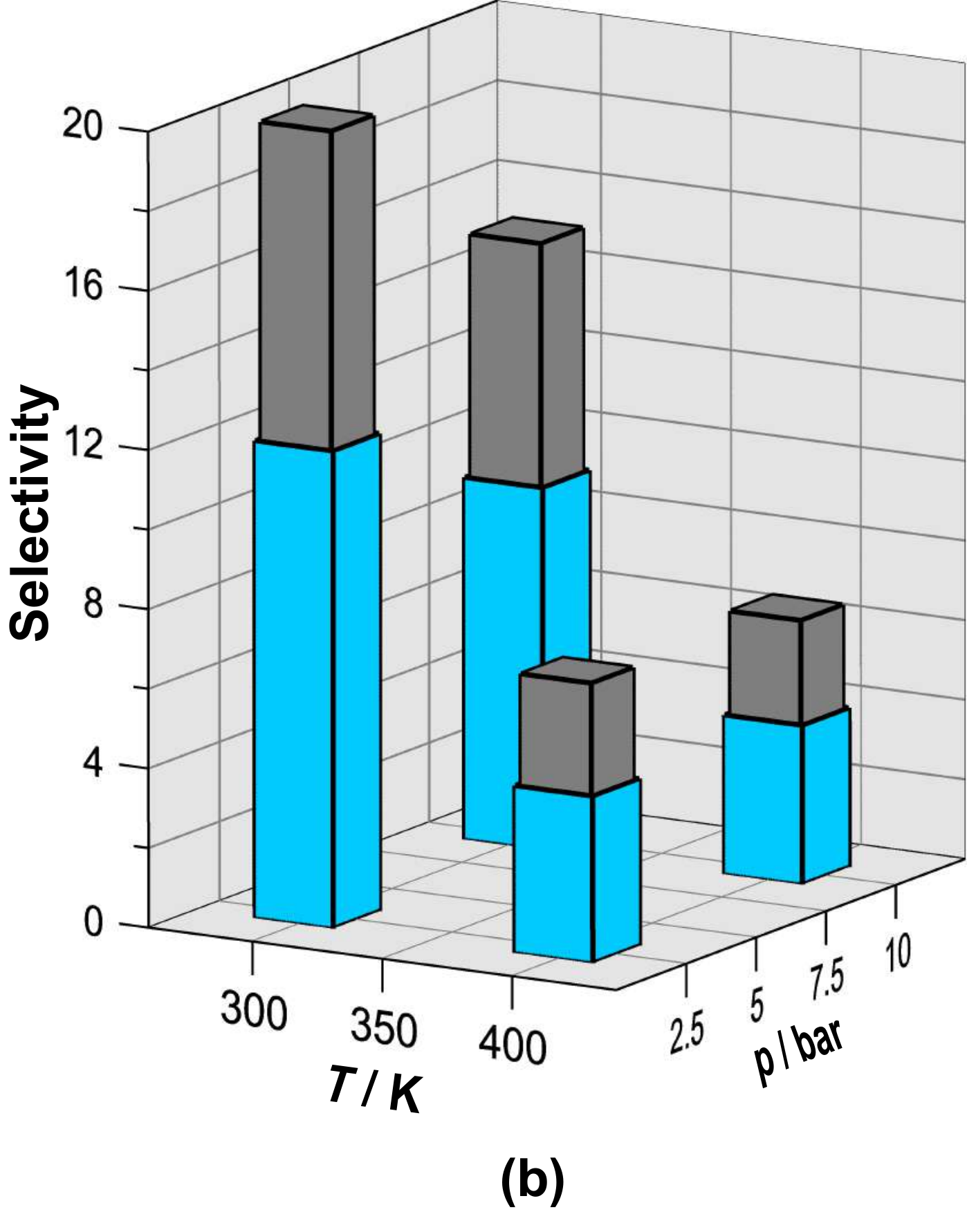}}\hspace{0.5cm}\scalebox{0.31}{\includegraphics{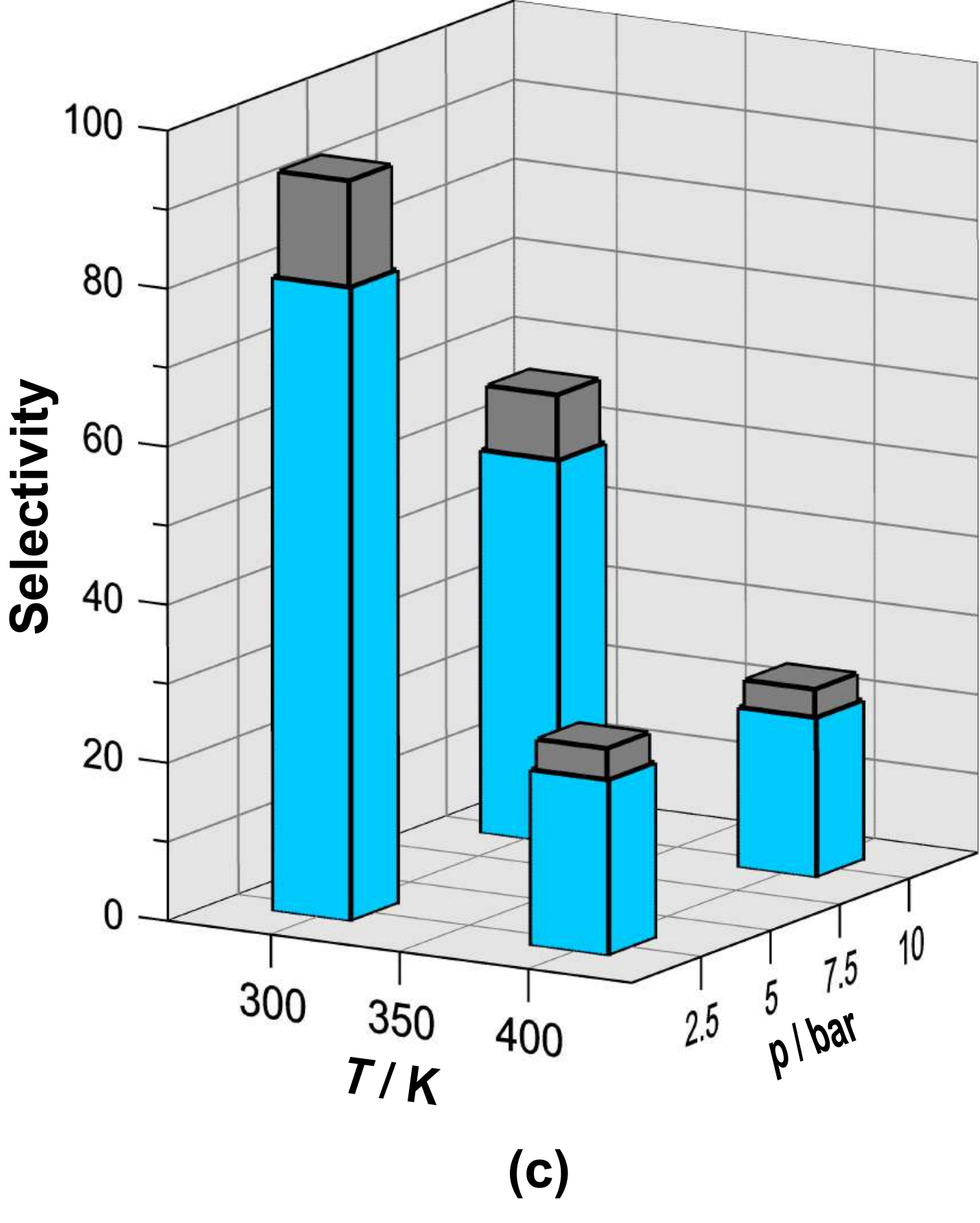}}
\end{center}
\caption{(Color online) Equilibrium selectivities for the MFI (a), LTA (b) and DDR (c) adsorbents with equimolar CH$_{4}$-H$_{2}$ gas mixture. Blue bars represent approximated selectivities using the corrected adsorption loadings due to the presence of inaccessible cavities. Note that the scales are different for each of the panels.}
\label{Fig4}
\end{figure}
\begin{figure}[!b]
\begin{center}
\scalebox{0.6}{\includegraphics{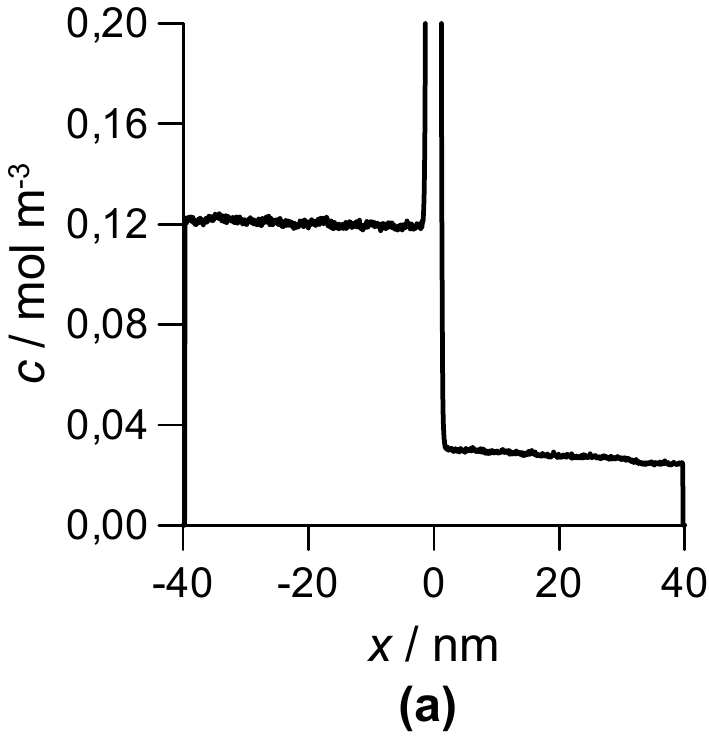}}\hspace{0.8cm}\scalebox{0.6}{\includegraphics{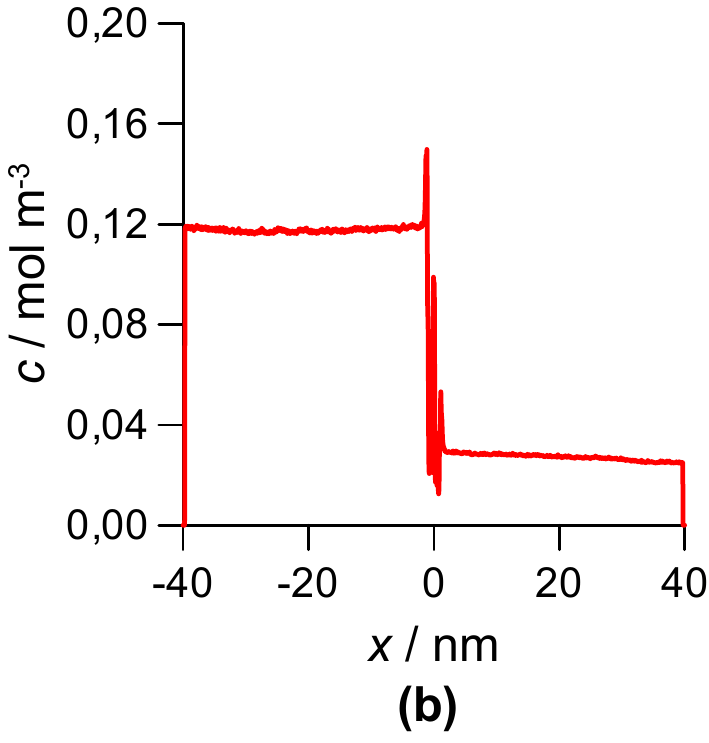}}\\ \vspace{0.5cm}
\scalebox{0.6}{\includegraphics{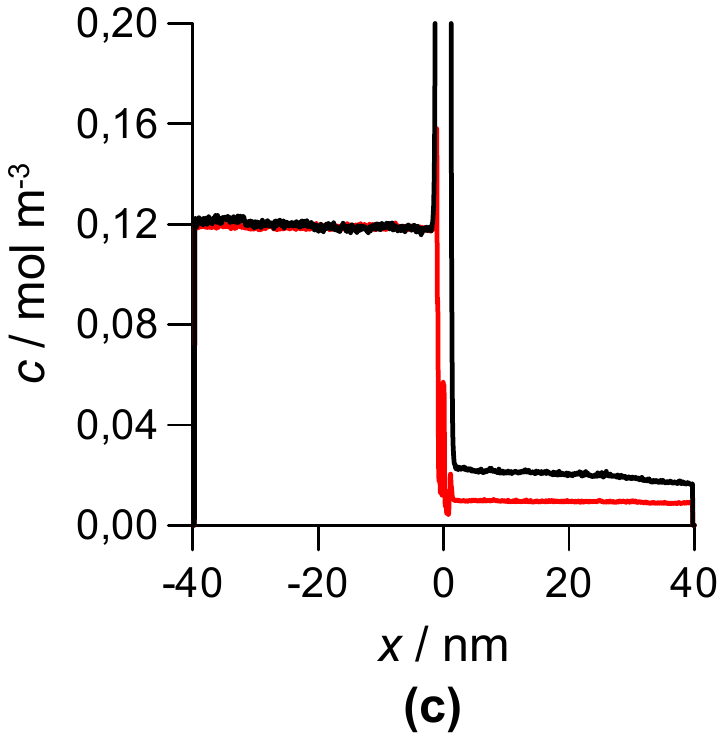}}\hspace{0.8cm}\scalebox{0.6}{\includegraphics{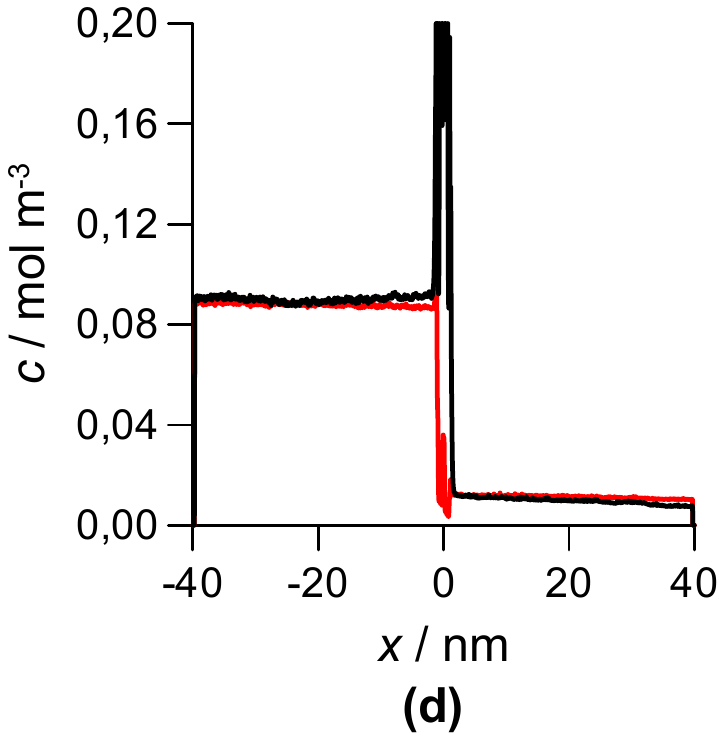}}
\end{center}
\caption{(Color online) Steady-state concentration profiles in the transport simulations with the MFI membrane (zig-zag channels): (a) pure CH$_{4}$ at 298.15~K and $p = 5 $~bar, (b) pure H$_{2}$ at 298.15~K and $p = 5$~bar, (c) CH$_{4}$-H$_{2}$ mixture at 298.15~K and $p = 10$~bar, (d) CH$_{4}$-H$_{2}$ mixture at 398.15~K and $p = 10$~bar (CH$_{4}$, black line; H$_{2}$, red line). Note that the concentration excesses of CH$_{4}$ in the membrane are only plotted partially for clarity purposes.}
\label{Fig5}
\end{figure}

Mixture adsorption was studied using GCMC simulations with an equimolar CH$_{4}$-H$_{2}$ gas system at temperatures of 298.15 and 398.15~K, and at the total pressures of 250 and 1000~kPa (more accurately, the CH$_{4}$-H$_{2}$ mixture was a nearly equimolar mixture, as we set the partial pressures of the components to be equal).
The obtained equilibrium loading and selectivity values are shown in figure~\ref{Fig4}.
The $S_{\mathrm{E}}$ values are between $\sim$4 and $\sim$80, so the degree of adsorption is noticeably higher for CH$_{4}$ than for H$_{2}$.
We found the absolute loadings with H$_{2}$ to be rather low (not shown), and this suggests some exclusion effect against the smaller gas molecules.
The relatively larger $S_{\mathrm{E}}$ values for the model DDR zeolite are due to its higher loadings with CH$_{4}$ (even if we consider the necessary corrections with the ratio of its inaccessible cavities).
Generally, the calculated selectivity decreases with an increasing temperature and pressure.

Our simulations using the PBD-MD technique delivered results for binary gas transport through the zeolite membranes with the CH$_{4}$ and H$_{2}$ gases at temperatures of 298.15 and 398.15~K, and at the total feed side pressures of 250 and 1000~kPa.
The permeate side pressure was always equal to 100~kPa (approximately ambient condition).
As a reference, single-component transport simulations were also performed, where the feed side pressure was equal to either the partial pressure of the component or to the total feed side pressure of the corresponding mixture simulation.

Before we proceed with the permeation results, we give a validation of our PBD-MD approach for the present conditions.
The concentration profile of the transporting particles is a good indicator whether the control cells can be considered as the ones representing bulk phases or not.
For a few simulations with the MFI zeolite, figure~\ref{Fig5} illustrates that there are practically constant concentrations along the $x$ direction in the control cells.
Slight deviations from the horizontal line (and thus bulk character) can only be detected near the boundary region of the permeate side.
In these diluted gases, the use of identical pressures or partial pressures on the feed side gives rise to very similar concentration profiles of the unlike components.
The obtained concentration ratio of systems of different temperatures also properly reflects the applied temperature ratio [figure~\ref{Fig5}~(c), (d)].
It should be noted, furthermore, that the calculated total pressure on the permeate side was always correct in the simulations (the target pressures were generally reproduced within 1\%), even if the partial pressures and component concentrations looked different here [e.g., figure~\ref{Fig5}~(c)].
This corresponds to the selectivity-related properties of the membrane.

\begin{figure}[!b]
\begin{center}
\scalebox{0.31}{\includegraphics{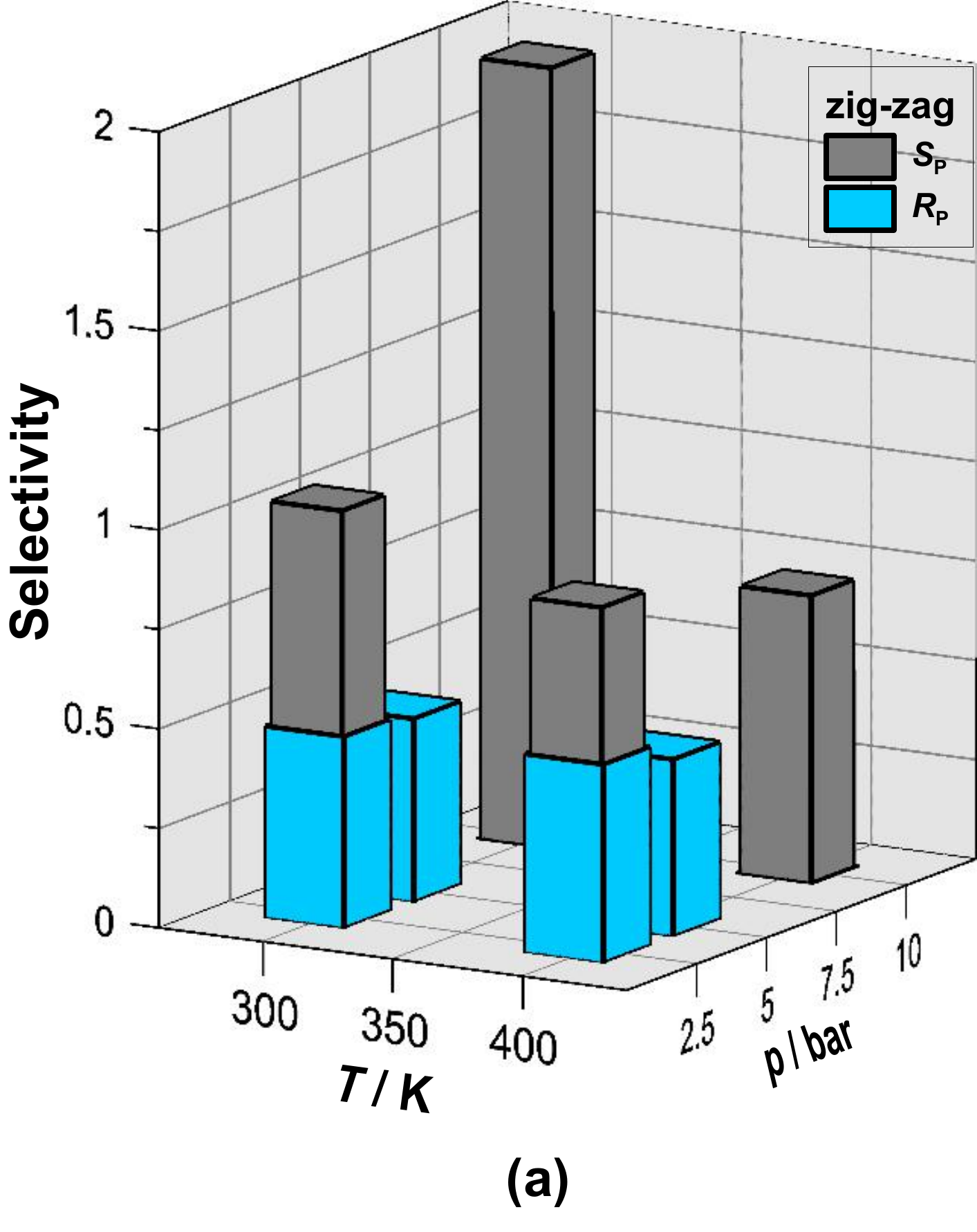}}\hspace{0.5cm}\scalebox{0.31}{\includegraphics{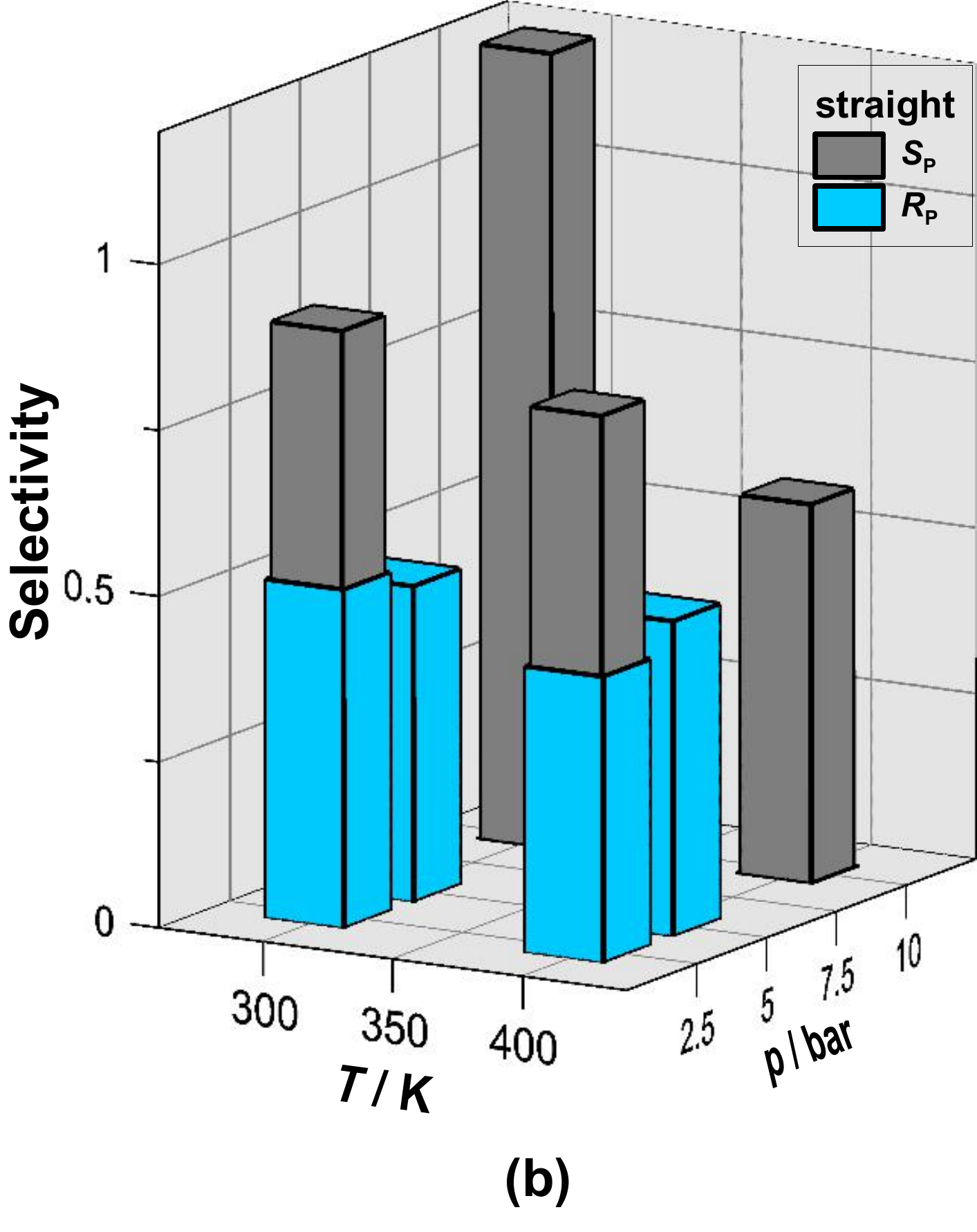}}\hspace{0.5cm}\scalebox{0.31}{\includegraphics{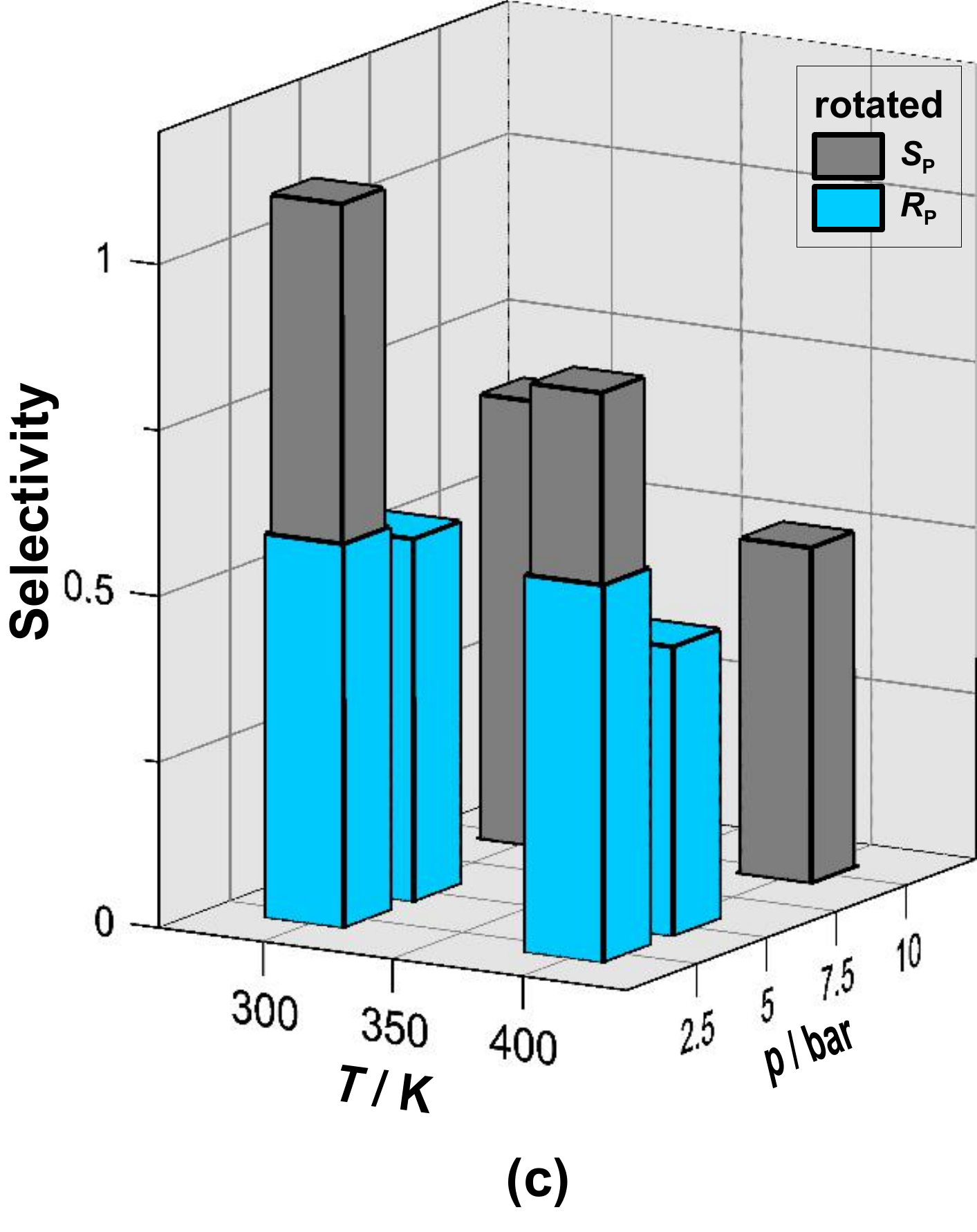}}\\ \vspace{0.3cm}
\scalebox{0.31}{\includegraphics{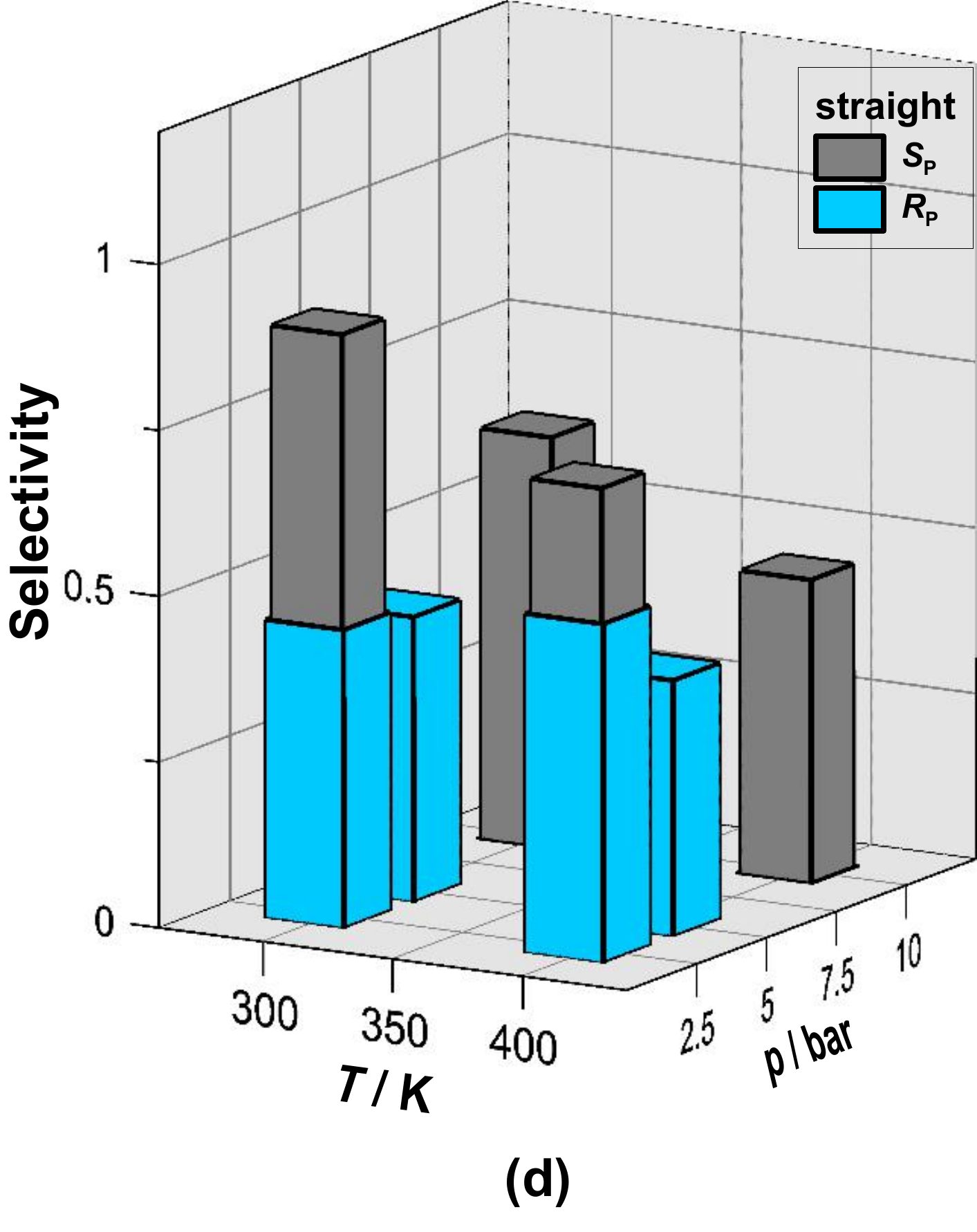}}\hspace{0.5cm}\scalebox{0.31}{\includegraphics{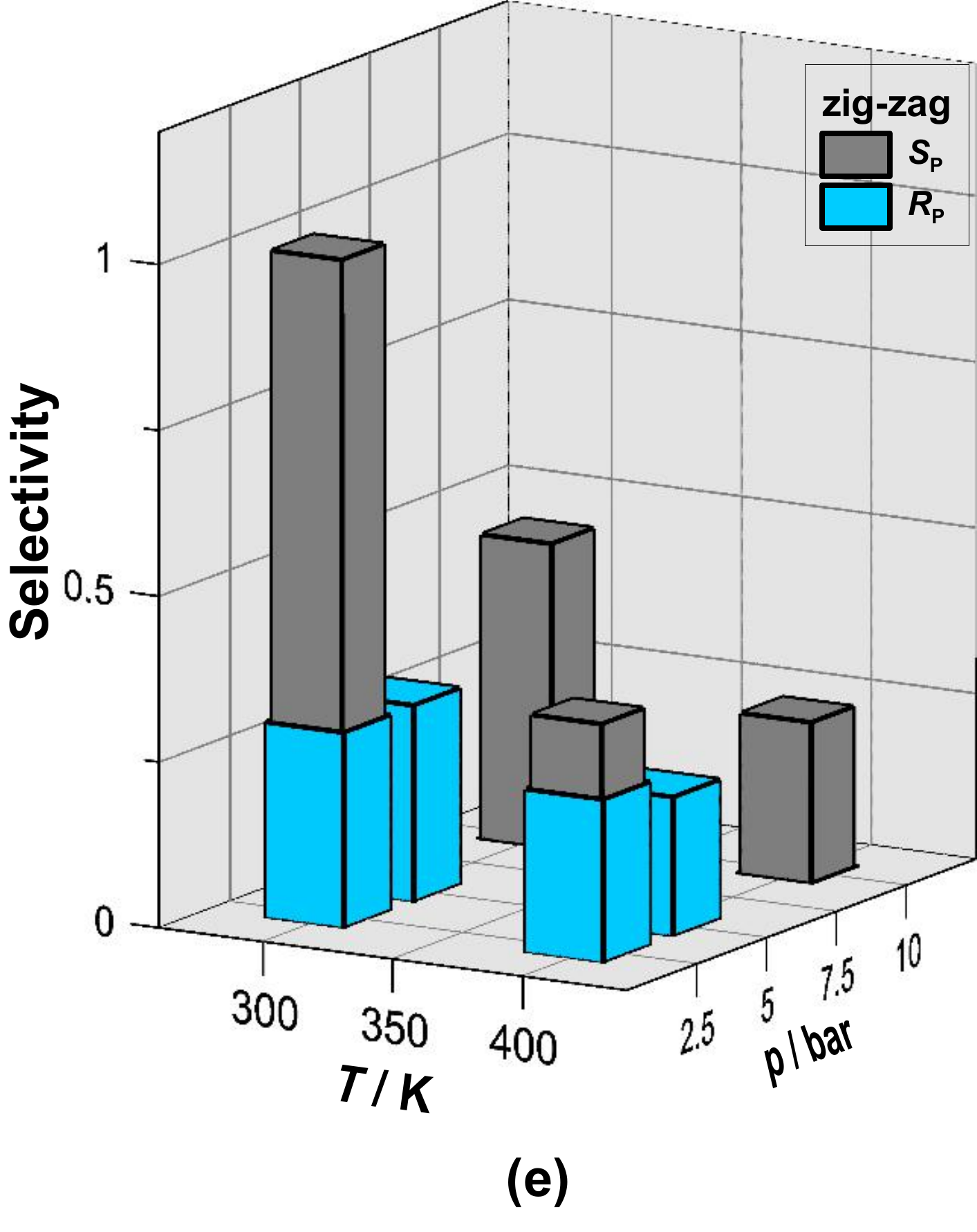}}\hspace{0.5cm}\scalebox{0.31}{\includegraphics{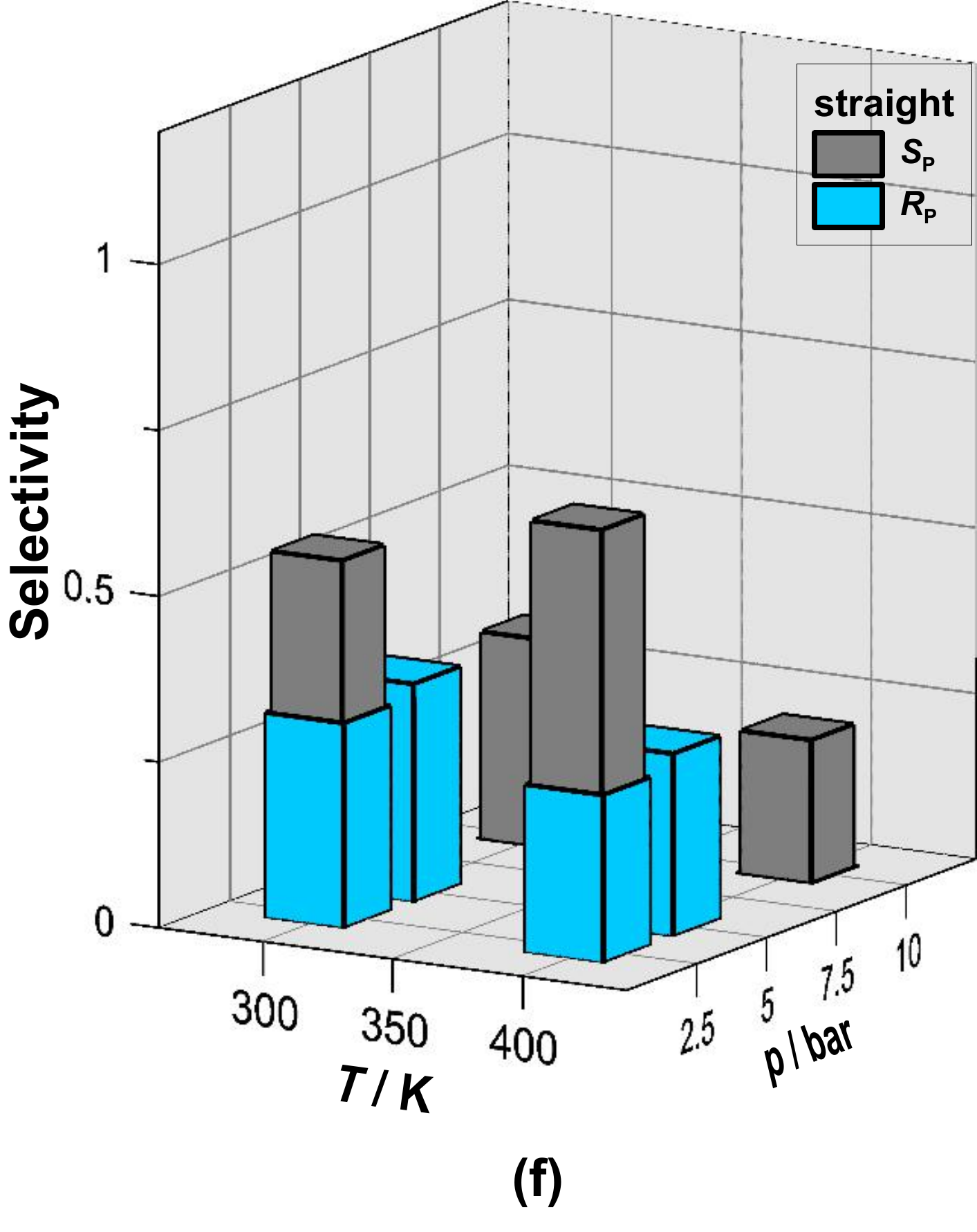}}
\end{center}
\caption{(Color online) Permeation selectivities (grey bars) and permeation ratios (blue bars) for the MFI [(a), (b)], LTA [(c), (d)] and DDR [(e), (f)] membranes with equimolar CH$_{4}$-H$_{2}$ gas mixture. Note that the plot for the MFI membrane with zig-zag channels (a) differs from the others in the scale.}
\label{Fig6}
\end{figure}

The calculated permeation selectivities and permeation ratios are depicted in figure~\ref{Fig6}.
It is seen that the $S_{\mathrm{P}}$ values are mostly less than 1, which means that, in their majority, the model membranes are selective for the transport of H$_{2}$.
Sufficiently high or low selectivity data for practical use are absent in the figure.
Separation of CH$_{4}$ from H$_{2}$ on the permeate side is only reliable with the zig-zag model MFI membrane at a lower temperature and higher pressure, where the dynamical selectivity is almost equal to 2.
On the other hand, separation of H$_{2}$ from CH$_{4}$ is preferred with the model DDR membranes, especially at a higher temperature and pressure ($S_{\mathrm{P}} < 0.25$).
In this respect, the model LTA membranes are of intermediate type.
The calculated $S_{\mathrm{P}}$ values are consistently larger than the corresponding $R_{\mathrm{P}}$ values, indicating the presence of momentum coupling between the unlike gas molecules during the mixture transport and, more specifically, the effect of the prevalent adsorption of CH$_{4}$.
The influence on each other's transport flux can also be recognized from the absolute values of the component fluxes because, generally, we obtained somewhat larger values for the pure component transport.
Similarly to the mixture transport data (see table~\ref{tab3}), the pure component fluxes (not shown) are the lowest for the model DDR membranes (the deviation is more significant for CH$_{4}$).
For the investigated membranes and conditions, the most remarkable fact is that the permeation selectivity values are much lower than the equilibrium selectivity values.
It means that the adsorption preference for CH$_{4}$ observed in all equilibrium cases does not predictably go hand in hand with transport selectivity in favor of CH$_{4}$.
This remains true even if we know that the feed side pressure in the dynamic simulations cannot be unequivocally compared to the equilibrium pressure of the adsorption simulations.
In the light of the pressure gradient applied in the transport simulations across the membranes, however, it is somewhat logical that the equilibrium adsorption loadings are generally larger than the dynamic loadings of the corresponding membranes with $p_{\mathrm{feed}} = p_{\mathrm{adsorption}}$ (compare data of table~\ref{tab3}).

\begin{table}[!t]
\centering
\caption{Equilibrium and dynamic concentrations ($c$) on the different zeolites [adsorbent (denoted by a) or membranes (denoted by m)] and transport fluxes ($j$) obtained with equimolar CH$_{4}$-H$_{2}$ gas mixture. }
\label{tab3}
\vspace{2ex}
\begin{tabular}{|c|cc|cc|cc|cc|}
\hline \hline
 & \multicolumn{4}{c|}{zig-zag (or rotated) channels} & \multicolumn{4}{c|}{straight channels} \\ \hline\hline
$T$ / K & \multicolumn{2}{c|}{298.15} & \multicolumn{2}{c|}{398.15} & \multicolumn{2}{c|}{298.15} & \multicolumn{2}{c|}{398.15} \\ \hline
$p$ / bar & 2.5 & 10 & 2.5 & 10 & 2.5 & 10 & 2.5 & 10 \\ \hline
$c_{\mathrm{CH}_{4}}$/(mol/dm$^{3}$ MFI$^\mathrm{a}$) & 1.255 & 2.887 & 0.224 & 0.787 & 1.255 & 2.887 & 0.224 & 0.787 \\
$c_{\mathrm{H_{2}}}$/(mol/dm$^{3}$ MFI$^\mathrm{a}$) & 0.027 & 0.065 & 0.016 & 0.058 & 0.027 & 0.065 & 0.016 & 0.058 \\
$c_{\mathrm{CH}_{4}}$/(mol/dm$^{3}$ MFI$^\mathrm{m}$) & 0.848 & 1.493 & 0.256 & 0.455 & 1.007 & 1.747 & 0.324 & 0.524 \\
$c_{\mathrm{H}_{2}}$/(mol/dm$^{3}$ MFI$^\mathrm{m}$) & 0.020 & 0.054 & 0.010 & 0.034 & 0.018 & 0.048 & 0.008 & 0.030 \\
$j_{\mathrm{CH}_{4}}$/(ns$\cdot$nm$^{2}$ MFI$^\mathrm{m}$) & 0.17 & 1.00 & 0.08 & 0.37 & 0.09 & 0.94 & 0.14 & 0.66 \\
$j_{\mathrm{H}_{2}}$/(ns nm$^{2}$ MFI$^\mathrm{m}$) & 0.16 & 0.51 & 0.09 & 0.51 & 0.10 & 0.78 & 0.17 & 1.15 \\ \hline
$c_{\mathrm{CH}_{4}}$/(mol/dm$^{3}$ LTA$^\mathrm{a}$)$^{\ast 0.6}$ & 0.47 & 1.19 & 0.10 & 0.34 & 0.47 & 1.19 & 0.10 & 0.34 \\
$c_{\mathrm{H}_{2}}$/(mol/dm$^{3}$ LTA$^\mathrm{a}$) & 0.039 & 0.131 & 0.023 & 0.087 & 0.039 & 0.131 & 0.023 & 0.087 \\
$c_{\mathrm{CH}_{4}}$/(mol/dm$^{3}$ LTA$^\mathrm{m}$) & 0.441 & 0.918 & 0.140 & 0.286 & 0.637 & 1.103 & 0.196 & 0.358 \\
$c_{\mathrm{H}_{2}}$/(mol/dm$^{3}$ LTA$^\mathrm{m}$) & 0.024 & 0.075 & 0.012 & 0.045 & 0.019 & 0.064 & 0.010 & 0.036 \\
$j_{\mathrm{CH}_{4}}$/(ns$\cdot$nm$^{2}$ LTA$^\mathrm{m}$) & 0.14 & 0.56 & 0.08 & 0.36 & 0.17 & 0.62 & 0.10 & 0.37 \\
$j_{\mathrm{H}_{2}}$/(ns$\cdot$nm$^{2}$ LTA$^\mathrm{m}$) & 0.15 & 0.90 & 0.11 & 0.80 & 0.15 & 0.92 & 0.12 & 0.74 \\ \hline
$c_{\mathrm{CH}_{4}}$/(mol/dm$^{3}$ DDR$^\mathrm{a}$)$^{\ast 0.6}$ & 1.58 & 2.43 & 0.31 & 0.93 & 1.58 & 2.43 & 0.31 & 0.93 \\
$c_{\mathrm{H}_{2}}$/(mol/dm$^{3}$ DDR$^\mathrm{a}$)$^{\ast 0.7}$ & 0.02 & 0.05 & 0.01 & 0.05 & 0.02 & 0.05 & 0.01 & 0.05 \\
$c_{\mathrm{CH}_{4}}$/(mol/dm$^{3}$ DDR$^\mathrm{m}$) & 0.808 & 1.484 & 0.294 & 0.435 & 0.953 & 1.486 & 0.296 & 0.473 \\
$c_{\mathrm{H}_{2}}$/(mol/dm$^{3}$ DDR$^\mathrm{m}$) & 0.016 & 0.048 & 0.007 & 0.025 & 0.016 & 0.048 & 0.007 & 0.025 \\
$j_{\mathrm{CH}_{4}}$/(ns$\cdot$nm$^{2}$ DDR$^\mathrm{m}$) & 0.06 & 0.11 & 0.01 & 0.08 & 0.03 & 0.10 & 0.02 & 0.06 \\
$j_{\mathrm{H}_{2}}$/(ns$\cdot$nm$^{2}$ DDR$^\mathrm{m}$) & 0.06 & 0.34 & 0.04 & 0.35 & 0.06 & 0.35 & 0.03 & 0.30
 \\ \hline \hline
 \multicolumn{9}{l}{$^{\ast}$ due to inaccessible cavities, simulated data are corrected by the factor indicated.}
\end{tabular}
\end{table}

The $S_{\mathrm{P}}$ values predominantly exhibit some decrease with an increasing temperature, which is attributed to a smaller decrease of the flux of H$_{2}$ with temperature (cf.\ table~\ref{tab3}).
At the same time, the general declining pressure dependence of $S_{\mathrm{P}}$ observed here is due to a more intense increase of the flux of H$_{2}$ with pressure.
The only exception to this latter tendency is the transport through the model MFI membranes (with both straight and zig-zag channels) at a lower temperature, where the growth in the flux of CH$_{4}$ is larger.
This behavior cannot be explained by anomalous loadings of the MFI membrane in dynamic situations.
As data of table~\ref{tab3} show, the CH$_{4}$ to H$_{2}$ loading ratios of the membranes in the dynamic simulations roughly follow the trends of the $S_{\mathrm{E}}$ values obtained from the equilibrium adsorption simulations.
However, the much milder temperature and pressure dependences of the $R_{\mathrm{P}}$ values than those of the $S_{\mathrm{P}}$ values (see figure~\ref{Fig6}) underline the importance of the momentum coupling between different components connected to some adsorption preferences.

\section{Discussion and conclusions}
\label{sec:conc}

We used the novel PBD-MD simulation technique to study the stationary membrane transport at the molecular level.
The method makes possible an accurate determination of steady-state fluxes of gases through microporous membranes.
We investigated the permeation of CH$_{4}$ and H$_{2}$ gases across pure silica zeolite membranes (MFI, LTA and DDR types) at 25\degree{C} and 125\degree{C} and at 2.5 and 5 (or 10)~bar.

The applied shifted and cut LJ potential is a less detailed interaction model for the simulations, and certainly, we could have better reproduced the available experimental data for the adsorption of pure CH$_{4}$ and H$_{2}$ gases by using more realistic all-atom potential models involving sophisticated inter- and intramolecular interactions.
By this simple modelling, we rather preserved the transferability of the zeolite parameters and stressed the structural differences of the investigated zeolites.
On the other hand, in such a way the total computational need of the executed transport simulations could be kept at a relatively low level (a couple of months with some dozens of CPU cores in a cluster computer).

We did not find experimental data for the investigated transport systems.
At the same time, there would be many difficulties with the comparison of experimental and such atomistic simulation results.
Experimental (synthesized) zeolite membranes typically consist of microcrystals with random orientations, and consequently, they exhibit an anisotropic pore geometry.
The permeation properties of these membranes can be very different from those with oriented microcrystals \cite{32} (these latter can be prepared in exceptional synthesis conditions).
Moreover, real transport processes in zeolite membranes can occur in the intercrystalline pores.
Other common problems include the presence of defects in the microcrystals and contaminations on the membrane surfaces, which may significantly affect the flux and selectivity of the membranes in experiments.
Furthermore, an experimental zeolite membrane is built up from a thin zeolite layer and a macroporous supporting layer (this provides the membrane with the requisite mechanical strength); the support resistance may also influence the separation performance.
From the simulation viewpoint, the main obstacle is that the extremely high demand of computational power today prevents the atomistic simulation of systems with real-life size (or nearly real-life size) membrane thicknesses (which is in the order of a micrometer).
According to our former experience, the dependence of the transport properties on membrane thickness calculated with commonly applied simulation system sizes (with $1, 2, 3,\ldots,$ unit-cell thick membranes) cannot be easily used to forecast the magnitude of the scale-up effect, since this dependence is not necessarily linear at the atomic scale.
For these reasons, we consider our present calculations only as one step towards model studies of the interplay of adsorption and diffusion in the permeation processes with realistic microporous membrane structures.

\looseness=-1 The results for the investigated systems and conditions have two striking features.
First, there are no profound differences between the three microporous zeolite structures of similar channel diameter concerning both their equilibrium and transport selectivities.
Second, the most important trend that is revealed from the present molecular simulations is that these model zeolites suffer very low transport selectivities for CH$_{4}$ over H$_{2}$ as compared to the equilibrium ones.
Higher equilibrium adsorption of CH$_{4}$ observed with these zeolites is more frequently linked with the permeation selectivity for H$_{2}$ over CH$_{4}$.
The contrast between the equilibrium and transport selectivity data can be ascribed to the quite different adsorption preferences of the gas molecules, the strongly dissimilar mobilities of the components in the membrane and their dependence on the loading of the membrane.
Here, a lower mobility of one gas component is compensated by a more intense adsorption capability, and the reverse situation occurs with the other component, leading to very similar calculated component fluxes, and consequently, to permeation selectivity values not far from 1.
The manifestation of this compensating effect is most remarkable in the case of the model DDR zeolite, where one of the largest equilibrium loadings of the adsorbent is combined with the significantly lower fluxes of the membranes.
A certain role of the different degree of temporary sorption of molecules in dynamic conditions is evidenced by the fact that for each zeolite, somewhat higher permeation selectivities were obtained for the zig-zag channels of longer transfer route than for the straight channels, which have an otherwise similar diameter.
Note that the applied pair of the LTA model membranes is a special case, where the transfer route of one of these membranes is artificially lengthened through rotating the straight channels by 45\degree.
At the same time, the latter observation with the LTA membranes suggests that the use of larger membrane thicknesses would probably result in the increase of permeation selectivity.
We obtained a truly higher permeation rate of CH$_{4}$ in only one case: for the model MFI membrane with zig-zag channels in the direction of the transport at a lower temperature and higher pressure.
This prevalent transport of CH$_{4}$ through the membrane seems to be a favorable interplay between a relatively high equilibrium selectivity and a smaller channel diameter with an optimal channel structure/orientation, which latter impedes the diffusion of H$_{2}$ molecules to a greater extent.
In the present systems, the transport through the membranes was more often easier for the far faster moving and smaller H$_{2}$ molecules.
To this end, however, there was one condition to be fulfilled: the absence of a significant steric hindrance from the adsorbed or slowly moving CH$_{4}$ molecules. In our case, apparently, the mobility of the H$_{2}$ molecules was not noticeably reduced due to such a blocking effect.

\section*{Acknowledgements}

Present article was published within the framework of the project GINOP-2.3.2-15-2016-00053 [``Development of engine fuels with high hydrogen content in their molecular structures (contribution to sustainable mobility)''].
We gratefully acknowledge the support of the Hungarian National Research Fund (OTKA NN113527) in the framework of ERA Chemistry, too.

\ukrainianpart
\title{Дослідження  мембранного відокремлювання для сумішей газів метан-водень }

\author{Т. Ковач, С. Папп, Т. Крістоф}
\address{Факультет фізичної хімії, університет Паннонії, м. Веспрем, H-8201, Угорщина
 }

\makeukrtitle

\begin{abstract}

Представлено результати прямих симуляцій для транспорту стаціонарного газу  крізь чисті кремнеземові цеолітові мембрани (MFI, LTA і DDR типу), використовуючи недавно запропоновану симуляційну методологію гібридної нерівноважної молекулярної динаміки.
Моделі міжмолекулярних потенціалів для газів CH$_{4}$ і H$_{2}$, що вивчалися, було взято з літератури.
Для різних цеолітів використано однакові параметри атомної (Si і O) взаємодії,  а мембрани конструювалися відповідно до їхніх реальних  (MFI, LTA чи DDR) кристалічних структур.
Реалістичну природу застосованих параметрів потенціалів протестовано шляхом здійснення симуляцій рівноважної адсорбції і порівняння обчислених результатів з експериментальними даними для ізотерм адсорбції.
Результати симуляцій транспорту, отримані при 25\degree{C} і 125\degree{C} і при 2.5, 5 чи 10~бар, чітко показують, що  селективності проникнення  CH$_{4}$ є вищими, ніж відповідні коефіцієнти проникнення чистих компонент, і значно відрізняються від рівноважних селективностей в  адсорбції сумішей. Селективність переносу на користь  CH$_{4}$ була спостережена тільки в одному випадку. Велика розбіжність між двома типами даних для селективності може бути атрибутом неподібності в рухливості компонент в мембрані, їхньої залежності від завантаження мембрани, а також відмінностей в адсорбційних преференціях молекул газу.

\keywords проникнення газу, цеолітова мембрана, стаціонарний режим, молекулярна динаміка
\end{abstract}

\begin{thebibliography}{99}

\bibitem{1} Fetting F., Chem. Ing. Tech., 1992, \textbf{64}, No. 3, 288, \doi{10.1002/cite.330640323}.

\bibitem{2} Auerbach S., Carrado K., Dutta P. (Eds.), Handbook of Zeolite Science and Technology, CRC Press, New York, 2003, \doi{10.1201/9780203911167}.

\bibitem{3} Nishiyama N., Yamaguchi M., Katayama T., Hirota Y., Miyamoto M., Egashira Y., Ueyama K.,
Nakanishi K., Ohta T., Mizusawa A., Satoh T., J. Membr. Sci., 2007, \textbf{306}, No. 1--2, 349, \doi{10.1016/j.memsci.2007.09.011}.

\bibitem{4} Mitchell M.C., Autry J.D., Nenoff T.M., Mol. Phys., 2001, \textbf{99}, No. 22, 1831, \doi{10.1080/00268970110075752}.

\bibitem{5} Heffelfinger G.S., van Swol F., J. Chem. Phys., 1994, \textbf{100}, No. 10, 7548, \doi{10.1063/1.466849}.

\bibitem{6} Im W., Seefeld S., Roux B., Biophys. J., 2000, \textbf{79}, No. 2, 788, \doi{10.1016/S0006-3495(00)76336-3}.

\bibitem{7} L\'isal M., Brennan J.K., Smith W.R., Siperstein F.R., J. Chem. Phys., 2004, \textbf{121}, No. 10, 4901,\\ \doi{10.1063/1.1782031}.

\bibitem{8} Rutkai G., Krist\'of T., J. Chem. Phys., 2010, \textbf{132}, No. 10, 104107, \doi{10.1063/1.3359434}.

\bibitem{9} Rutkai G., Boda D., Krist\'of T., J. Phys. Chem. Lett., 2010, \textbf{1}, No. 23, 2179, \doi{10.1021/jz100718n}.

\bibitem{10} Cs\'anyi \'E., Boda D., Gillespie D., Krist\'of T., Biochim. Biophys. Acta, Biomembr., 2012,
\textbf{1818}, No. 3, 592, \doi{10.1016/j.bbamem.2011.10.029}.

\bibitem{11} Boda D., Gillespie D., J. Chem. Theory Comput., 2012, \textbf{8}, No. 3, 824, \doi{10.1021/ct2007988}.

\bibitem{12} Hat\'o Z., Boda D., Krist\'of T., J. Chem. Phys., 2012, \textbf{137}, No. 5, 054109, \doi{10.1063/1.4739255}.

\bibitem{13} Maginn E.J., Bell A.T., Theodorou D.N., J. Phys. Chem., 1993, \textbf{97}, No. 16, 4173, \doi{10.1021/j100118a038}.

\bibitem{14} Kjelstrup S., Bedeaux D., Inzoli I., Simon J.M., Energy, 2008, \textbf{33}, No. 8, 1185, \doi{10.1016/j.energy.2008.04.005}.

\bibitem{15} Xu J., Kjelstrup S., Bedeaux D., Phys. Chem. Chem. Phys., 2006, \textbf{8}, No. 17, 2017, \doi{10.1039/B516704C}.

\bibitem{16} Huang C., Nandakumar K., Choi P.Y.K., Kostiuk L.W., J. Chem. Phys., 2006, \textbf{124}, No. 23, 234701, \doi{10.1063/1.2209236}.

\bibitem{17} Evans D.J., Morriss O., Comput. Phys. Rep., 1984, \textbf{1}, No. 6, 297, \doi{10.1016/0167-7977(84)90001-7}.

\bibitem{18} Frentrup H., Avenda\~no C., Horsch M., Salih A., M\"uller E.A., Mol. Simul., 2012, \textbf{38}, No. 7, 540, \doi{10.1080/08927022.2011.636813}.

\bibitem{19} Hat\'o Z., Kaviczki \'A., Krist\'of T., Mol. Simul., 2016, \textbf{42}, No. 1, 71, \doi{10.1080/08927022.2015.1010083}.

\bibitem{20} Berti C., Furini S., Gillespie D., J. Chem. Theory Comput., 2016, \textbf{12}, No. 3, 925, \doi{10.1021/acs.jctc.5b01044}.

\bibitem{21} Database of zeolite structures, URL~\url{http://www.iza-structure.org/databases}.

\bibitem{22} Demontis P., Fois E.S., Suffritti G.B., Quartieri S., J. Phys. Chem., 1990, \textbf{94}, No. 10, 4329,\\ \doi{10.1021/j100373a083}.

\bibitem{23} Skoulidas A.I., Sholl D.S., J. Phys. Chem. B, 2002, \textbf{106}, No. 19, 5058, \doi{10.1021/jp014279x}.

\bibitem{24} Michels A., de Graaff W., Seldam C.A.T., Physica, 1960, \textbf{26}, No. 6, 393, \doi{10.1016/0031-8914(60)90029-X}.

\bibitem{25} Van den Berg A.W.C., Bromley S.T., Jansen J.C., Microporous Mesoporous Mater., 2005, \textbf{78}, No. 1, 63, \doi{10.1016/j.micromeso.2004.09.017}.

\bibitem{26} Wirawan S.K., Petersson M., Creaser D., ASEAN J. Chem. Eng., 2008, \textbf{8}, No. 1, 9.

\bibitem{27} Jhung S.H., Yoon J.W., Lee J.S., Chang J.S., Chem. Eur. J., 2007, \textbf{13}, No. 22, 6502,\\ \doi{10.1002/chem.200700148}.

\bibitem{28} Palomino M., Corma A., Rey F., Valencia S., Langmuir, 2010, \textbf{26}, No. 3, 1910, \doi{10.1021/la9026656}.

\bibitem{29} Huth A.J., Stueve J.M., Guliants V.V., J. Membr. Sci., 2012, \textbf{403--404}, 236, \doi{10.1016/j.memsci.2012.02.056}.

\bibitem{30} Foster M.D., Rivin I., Treacy M.M.J., Friedrichs O.D., Microporous Mesoporous Mater., 2006, \textbf{90}, No. 1--3, 32, \doi{10.1016/j.micromeso.2005.08.025}.

\bibitem{31} First E.L., Gounaris C.E., Wei J., Floudas C.A., Phys. Chem. Chem. Phys., 2011, \textbf{13}, No. 38, 17339, \doi{10.1039/C1CP21731C}.

\bibitem{32} Lovallo M.C., Tsapatsis M., AlChE J., 1996, \textbf{42}, No. 11, 3020, \doi{10.1002/aic.690421104}.
\end{thebibliography}
\end{document}